\documentclass[a4paper,11pt,preprintnumbers]{article}
\pdfoutput=1

\usepackage{amssymb,mathrsfs,amsmath}
\usepackage{a4wide}
\usepackage{color,xcolor}
\usepackage{slashed,soul}
\usepackage{graphicx}
\usepackage{amsfonts}
\usepackage{lscape}
\def\linkcolor{cyan!70!black}

\usepackage[
colorlinks=true
,urlcolor=\linkcolor
,anchorcolor=\linkcolor
,citecolor=\linkcolor
,filecolor=\linkcolor
,linkcolor=\linkcolor
,menucolor=\linkcolor
,linktocpage=true
,pdfproducer=medialab
,pdfa=true
]{hyperref}
\usepackage{amsthm}
\usepackage{booktabs}
\usepackage{array}
\usepackage{rotating}
\usepackage[numbers, sort&compress]{natbib}
\usepackage{multirow}
\usepackage{float}
\usepackage[utf8]{inputenc}
\usepackage[T1]{fontenc}
\usepackage{appendix}
\usepackage{epsfig}
\usepackage{orcidlink}
\usepackage{comment}
\usepackage{upgreek}
\usepackage{braket}

\newcommand{\beq}{\begin{equation}} 
\newcommand{\eeq}{\end{equation}} 
\newcommand{\ba}{\begin{array}}  
\newcommand{\ea}{\end{array}} 
\newcommand{\bea}{\begin{eqnarray}}  
\newcommand{\eea}{\end{eqnarray} }  
\newcommand{\bal}{\begin{align}}
\newcommand{\eal}{\end{align}}   
\newcommand{\bi}{\begin{itemize}}  
\newcommand{\ei}{\end{itemize}}  
\newcommand{\ben}{\begin{enumerate}}  
\newcommand{\een}{\end{enumerate}}  
\newcommand{\bc}{\begin{center}}
\newcommand{\ec}{\end{center}} 
\newcommand{\bt}{\begin{table}}
\newcommand{\et}{\end{table}}  
\newcommand{\btb}{\begin{tabular}}
\newcommand{\etb}{\end{tabular}}

\def\arrvline{\hfil\kern\arraycolsep\vline\kern-\arraycolsep\hfilneg}

\newcommand{\cnb}{\text{C$\nu$B}}

\allowdisplaybreaks

\let\OLDthebibliography\thebibliography
\renewcommand\thebibliography[1]{
  \OLDthebibliography{#1}
  \setlength{\parskip}{0pt}
  \setlength{\itemsep}{0pt plus 0.3ex}
}

\begin{document}

\vspace{1cm}

\begin{titlepage}

\begin{flushright}
IFT-UAM/CSIC-25-38\\
 \end{flushright}
\vspace{0.2truecm}

\begin{center}
\renewcommand{\baselinestretch}{1.8}\normalsize
\boldmath
{\LARGE\textbf{
Cosmogenic neutrinos as probes of new physics
}}
\unboldmath
\end{center}

\vspace{0.4truecm}

\renewcommand*{\thefootnote}{\fnsymbol{footnote}}

\begin{center}

{
Luighi P. S. Leal$^1$\footnote{\href{mailto:luighi.leal@usp.br}{luighi.leal@usp.br}},
Daniel Naredo-Tuero$^2$\footnote{\href{mailto:daniel.naredo@ift.csic.es}{daniel.naredo@ift.csic.es}},
Renata Zukanovich Funchal$^1$\footnote{\href{mailto:zukanov@if.usp.br}{zukanov@if.usp.br}}
}

\vspace{0.7truecm}

{\footnotesize
$^1$Instituto de F\'{\i}sica, Universidade de S\~{a}o Paulo, C.P. 66.318, 05315-970 S\~{a}o Paulo, Brazil
\\[.5ex]
$^2$Departamento de F\'{\i}sica Te\'orica and Instituto de F\'{\i}sica Te\'orica UAM/CSIC,\\
Universidad Aut\'onoma de Madrid, Cantoblanco, 28049 Madrid, Spain
}

\vspace*{2mm}
%\today
\end{center}

\renewcommand*{\thefootnote}{\arabic{footnote}}
\setcounter{footnote}{0}

%\vspace{0.3cm}
\begin{abstract}
The scattering of extremely energetic cosmic rays with both cosmic microwave background and extragalactic background light, can produce $\mathcal{O}(10^{18} \,{\rm eV})$ neutrinos, known as cosmogenic neutrinos. These neutrinos are the only messengers from the extreme cosmic accelerators that can reveal the origin of the most energetic cosmic rays. Consequently, much effort is being devoted to achieving their detection. In particular, the GRAND project aims to observe the $\nu_\tau$ and  $\bar \nu_\tau$ components of the cosmogenic neutrino flux in the near future using radio antennas. In this work, we investigate how the detection of cosmogenic neutrinos by GRAND can be used to probe beyond the Standard Model physics. We identify three well-motivated scenarios which induce distinct features in the cosmogenic neutrino spectrum at Earth: neutrino self-interactions mediated by a light scalar ($\nu$SI), pseudo-Dirac neutrinos (PD$\nu$) and 
neutrinos scattering on ultra-light Dark Matter ($\nu$DM). We show these scenarios  can be tested by GRAND, using 10 years of cosmogenic neutrino data, in a region of parameter space complementary to current experiments. For the $\nu$SI model,, we find that GRAND can constrain the coupling to $\nu_\tau$ in the range [$10^{-2},10^{-1}$] for a scalar with mass in the range 0.1 to 1 GeV. For PD$\nu$, we find that GRAND is sensitive to sterile-active mass squared splitting in the range [$10^{-15},10^{-13}$] ${\rm eV}^2$. Finally, for the $\nu$DM model, assuming a heavy mediator, GRAND can do substantially better than the current limits from other available data. These results rely on the fact that the actual cosmogenic flux is around the corner, not far from the current IceCube limit.

\end{abstract}

\end{titlepage}

\tableofcontents

\section{Introduction}
\label{sec:intro}
Cosmic rays, since their discovery by Victor Hess in 1912, have been connected to advances in our  understanding of nature. It was in the study of cosmic rays that in the first half of the last century we observed the positron, the muon and the pion for the first time.   
In 1998 neutrino flavor oscillations 
were first demonstrated by observing the flavor composition of atmospheric neutrinos that arrived at the Earth and were produced by cosmic rays~\cite{Super-Kamiokande:1998kpq}. Since 2013 the IceCube Neutrino Observatory is  detecting an unexpected flux of very high-energy neutrinos (in the hundreds of TeV to tens of PeV range) of extragalactic origin~\cite{IceCube:2013low}. 
If this neutrino flux is related or not to the sources of high-energy cosmic rays is still an open question.
The characterization of the neutrino spectrum and flavor composition  beyond the current energy range attainable by IceCube will be crucial in resolving the sources. This may also provide new ways to probe the most extreme particle accelerators in our Universe.

These cosmic accelerators remain hidden ultimately because the Universe becomes opaque at scales $\sim 100$ Mpc 
for extreme energy cosmic rays (EECRs), with energies $\gtrsim$~50 EeV. 
This is because, at those energies, the scattering of EECRs
with photons of the cosmic microwave background (CMB) and of the extragalactic background light (EBL) triggers the $\Delta^+$ resonance~\cite{Berezinsky:1970xj,Stecker:1973sy,Hill:1983xs}. 
As a by-product of these resonant scatterings, the so-called cosmogenic neutrinos are produced  mainly through pion or neutron decays. Unstable nuclei, 
which result from photodisintegration or photo-pion production, can also suffer beta decay further adding to this cosmogenic neutrino flux~\cite{AlvesBatista:2018zui}.
Cosmogenic neutrinos can then carry information from cosmic accelerators located at much larger distances from Earth than the ones we can inspect with EECRs. 
Moreover,  cosmogenic neutrinos are expected to have energies of about 10$^{18}$
eV (1~EeV) and to be viable probes of the sources of EECRs since their imperviousness to magnetic fields and extremely low interaction rate imply that they arrive at Earth only modified 
by energy redshift and flavor oscillation.

Although the existence of cosmogenic neutrinos is basically guaranteed~\footnote{The first neutrino event observed by KM3NeT, KM3-230213A~\cite{KM3NeT:2025npi}, may be of cosmogenic origin.}, the theoretical prediction of the expected diffuse flux, however, is plagued by astrophysical uncertainties concerning the primary source properties.
When modeling this flux, assumptions have to be made about ingredients such as the chemical composition, the injected spectral index, the luminosity, the maximal energy attainable, and  the cosmological evolution. The diffuse flux  may also be the result of a homogeneous distribution of identical sources or of a mixture of different types of sources.
Fortunately, these astrophysical uncertainties  seem to mainly affect the overall normalization of the flux, which typically spans several orders of magnitude but has the same general spectral shape predicted by different astrophysical models~\cite{Ahlers:2012rz,Aloisio:2015ega,AlvesBatista:2018zui,Ackermann:2022rqc}. Note, however, that we might receive in this energy range also neutrinos directly from the sources, causing unexpected features in the spectrum. We assume here that the cosmogenic component is dominant throughout the energy range we analyze.

On the other hand, cosmogenic neutrinos may turn out to be invaluable not only for the investigations of these extreme astronomical sources, but also for probing  neutrino properties~\cite{Denton:2020jft,Valera:2022ylt,Ackermann:2022rqc} and testing specific Beyond the Standard Model (BSM) scenarios~\cite{Huang:2021mki,Huang:2022pce,GarciaSoto:2022vlw,Heighton:2023qpg,Kirk:2023fin}. 

There are a few next generation 
projects aiming to provide sensitivity for neutrinos with energy $E_\nu \sim 1$ EeV at an energy flux level $E_\nu^2 \, \Phi_\nu  \lesssim 10^{-9}$  GeV cm$^{-2}$ s$^{-1}$ sr$^{-1}$
 exploring radio detection techniques
currently developed by experiments such as ANITA~\cite{ANITA:2021xxh}, ARA~\cite{ARA:2019wcf} and ARIANNA~\cite{ARIANNA:2019scz} to measure radio pulses generated by charged particle cascades 
(Askaryan  emission) induced by neutrino interactions in matter.
The most promising are the 
Giant Radio Array for Neutrino Detection (GRAND)~\cite{GRAND:2018iaj} and the IceCube-Gen2 radio array~\cite{IceCube-Gen2:2021rkf}~\footnote{
There is also the Probe Of Extreme Multi-Messenger Astrophysics (POEMMA)~\cite{POEMMA:2020ykm} 
which is a NASA Astrophysics probe-class mission to identify  cosmogenic tau neutrinos above 20 PeV.}. The GRAND antennas  
are designed to detect inclined extensive air-shower (EAS)~\footnote{IceCube-Gen2, on the other hand, relies on detecting the radio waves from scatterings of neutrinos \textit{inside} the ice. Since they don't require the charged lepton to exit the Earth and produce a EAS, IceCube-Gen2 is approximately sensitive to all flavors. See, for example, Section 4 of  Ref.~\cite{IceCube-Gen2:2020qha}.}. 
This makes them mainly sensitive to tau neutrinos ($\nu_\tau$).
Cosmogenic  $\nu_\tau$'s can produce EAS emerging from the Earth (rock or ice) if  they interact producing a $\tau$ lepton close enough to the Earth's surface. Typically, at the expected energies, the 
$\tau$ lepton can travel a few tens of km
in matter before exiting in the atmosphere and decaying, thus generating an Earth-skimming EAS. 

There are two caveats 
we would like to point out here. First, neutrino-nucleon SM cross section predictions are accurate only up to $~10^{-1}$ EeV mainly due to  limitations of extrapolating parton distribution functions  at lower Bjorken-$x$ than currently accessible by 
colliders and fixed target accelerator experiments
~\cite{Connolly:2011vc,
Arguelles:2015wba, Garcia:2020jwr}. 
Second, the effect of final state radiation (FSR) 
has to be accounted for. It seems that if FSR is not 
included in the calculation it will lead to the underestimation of the parent neutrino energy by 
about 
5\% (smaller than the energy resolution of GRAND). But 
FSR also affects the 
regeneration process as it 
lowers the tau energy~\cite{Plestid:2024bva}.
The first problem can be 
solved in the future by means of experimental data at higher energies and possible improvements of theoretical 
techniques.
The second can be solved by incorporating FSR in the calculations. We estimate that together they can affect our results by 10-20\%.

The purpose of this paper is to assess the  use of these future radio arrays 
in investigating   BSM physics 
that can produce distinct  
features in the observable cosmogenic neutrino spectrum  (such as dips or bumps) that can be disentangled from the
astrophysical uncertainties in the 
normalization and shape of the 
standard flux.
We start in section~\ref{sec:flux} by describing how we model and simulate the cosmogenic neutrino flux.
In section~\ref{sec:detector}
 we describe how we model the GRAND radio array detector response to Earth-skimming $\nu_\tau$. Section~\ref{sec:BSM} is devoted to presenting the three BSM scenarios we will study here:  neutrino self-interactions, pseudo-Dirac neutrinos, and neutrinos scattering on ultra-light scalar Dark Matter. In section~\ref{sec:ana}
 we  introduce our analysis
 and examine our results. We draw our conclusions and give our final remarks in section~\ref{sec:conc}.
 We also provide two  appendices; in~\ref{app:transport}
we review the main ingredients behind the calculation of the cosmogenic  neutrino flux at the Earth and in~\ref{app:xsec_nuDM} we
give the expression for the neutrino-ultralight Dark Matter differential cross section.

\section{Modeling  of the Cosmogenic Neutrino Flux}
\label{sec:flux}

The EECRs spectrum and chemical composition is highly unknown, and there seems to be some incompatibility between different experiments. While Auger seems to disfavor a sizeable light component between 50 - 100 EeV~\cite{PierreAuger:2024flk}, the Telescope Array collaboration reports compatibility with a moderately light composition below 100 EeV~\cite{TelescopeArray:2024buq}. These data have been used to make predictions about the  diffuse cosmogenic neutrino flux~\cite{Romero-Wolf:2017xqe,AlvesBatista:2018zui,Heinze:2019jou}. In fact, global fits to the source composition of these EECR sources seem to disfavor a pure-proton composition~\cite{Ahlers:2017bV,PierreAuger:2024flk}.
However, the existence of a proton 
(or light nuclei) subdominant or even dominant component at the highest energies cannot be ruled out by data~\cite{PierreAuger:2022atd,TelescopeArray:2024buq,TelescopeArray:2024oux}.
If the recent  KM3NeT $\sim0.1$ EeV event~\cite{KM3NeT:2025npi} is indeed of cosmogenic origin, this may indicate a non-negligible proton component at those energies. 
In fact, cosmic-ray observations on Earth are dominated by sources at $\lesssim $ 300 Mpc and   sources located beyond this
distance could have a different behavior~\cite{Romero-Wolf:2017xqe}.
This component could be produced by a secondary population of sources
that could  exclusively accelerate protons to extreme energies or where heavier nuclei are disintegrated before escaping the source region.
This would have a limited effect on  the full ultra-high-energy cosmic ray spectrum. Even a  subdominant proton or light nuclei component would alter the expected cosmogenic neutrino flux since they produce significantly more neutrinos than heavier nuclei.

For the modeling and simulation of the sources of EECRs and the production of cosmogenic neutrinos, we closely follow Refs.~\cite{Moller:2018isk,vanVliet:2019nse}. In Appendix~\ref{app:transport} we give a brief description of the type of integro-differential equations one needs to solve to compute the flux $\Phi_{\nu}(E_\nu)$ at the Earth.
For the sake of simplicity, we 
will assume here pure-proton sources with a power-law injection spectrum 
\begin{equation}
      \frac{dN_p}{dE_p}\propto E_p^{-\gamma} \exp\left(-\frac{E_p}{E_{\rm max}} \right)\,,
      \label{eq:proton_flux}
\end{equation}
with the spectral index $\gamma$ and the proton energy $E_p$ varying in the range from 10 EeV to $E_{\rm max}$. Although we will not include it in our main analysis, for illustrative purposes, we will also consider in this section source compositions with an iron fraction $\alpha_S$,  characterized by the following spectral distribution:
\begin{equation}
    \frac{dN}{dE}=\left(1-\alpha_S\right)\dfrac{dN_p}{dE_p}+\alpha_S\dfrac{dN_{\rm Fe}}{dE_{\rm Fe}}\,,
\end{equation}
where the iron flux is given by an analogous expression as in Eq.~\eqref{eq:proton_flux}:
\begin{equation}
    \frac{dN_{\rm Fe}}{dE_{\rm Fe}}\propto E_{\rm Fe}^{-\gamma}\exp\left(-\frac{E_{\rm Fe}}{ZR_{\rm max}}{}\right)\,, 
\end{equation}
where $Z$ is the atomic number and $R_{\rm max}$ is the maximum rigidity, chosen to be equal to the $E_{\rm max}$ in the proton flux of Eq.~(\ref{eq:proton_flux}).
For the sake of simplicity, we do not include intermediate elements in our simulations.

The assumption of pure-proton sources might seem too optimistic given the constraints on the mass composition of EECRs. However, it has been  shown that the source composition is degenerate with the redshift evolution of the sources~\cite{Moller:2018isk,vanVliet:2019nse} since, for a fixed spectral index, both modify the normalization of the flux. Moreover, this is not the most optimistic case in terms of cosmogenic neutrino production, as intermediate nuclei, such as He and N, can lead to larger fluxes (see, for example, Appendix A of~\cite{Moller:2018isk}).

In our scenario we consider  a distribution of  identical proton sources with the same luminosity which extend up to redshift    $z_{\text{max}}=7$. We
 parametrise the source evolution as

\begin{equation}
    \text{SE}(z,m)=
    \begin{cases}
        \left(1+z\right)^m & z<1,\\
        2^m & 1\leq z < 4,\\
2^m\left(\dfrac{1+z}{5}\right)^{-3.5}&4\leq z <7,\\
    \end{cases}
    \label{eq:source_evolution}
\end{equation}
where the choice $m=3$ for the source evolution parameter would correspond to a redshift evolution similar to the star formation rate~\cite{Yuksel:2008cu}, while $m=0$ would mimic the typical redshift evolution of BL-Lac models~\cite{Caccianiga:2001vu,Ajello:2013lka}.
The range of $m$ depends on the class of the sources~\cite{Moller:2018isk,vanVliet:2019nse}. 
So this parametrization allows us to approximate the redshift evolution of different classes of sources.

We simulate the propagation and scattering of the EECR protons with the  CRPropa3 code~\cite{AlvesBatista:2016vpy}. 
This code includes all relevant interactions for the calculation of the cosmogenic flux (photo-pion production, pair production, nuclear $\beta$ decay) 
and adiabatic energy losses due to the 
expansion of the Universe.

We perform an extragalactic 1D simulation (i.e. neglecting magnetic fields) keeping track of both protons and neutrinos. Magnetic-field effects are expected to increase the cosmogenic neutrino flux  by a factor of a few at $E_\nu \sim 1$ EeV~\cite{Wittkowski:2018giy}.
The photon fields against which the EECR protons will scatter are  
the CMB and the EBL, and we use the EBL model from Ref.~\cite{2012MNRAS.422.3189G}. According to Ref.~\cite{AlvesBatista:2019rhs}
the choice of the EBL model has a minor effect as the CMB is the dominant photon field for neutrino production in the  energy range of interest.

In the free-streaming case, the neutrino flux is simply

\begin{align}
    \Phi_\nu(E_\nu) = \frac{1}{4\pi} \int_0^{z_{\rm max}}{\rm d}z \frac{{\cal L}_\nu (z,(z+1)E_\nu)}{H(z)}\,,
\end{align}
where ${\cal L}_\nu(z,E)={\rm SE}(z,m) \, {\cal L}_\nu(0,E)$ and ${\cal L}_\nu(0,E) \propto dN_p/dE_p$.

As a means of normalising the total flux, we make use of the EECR spectrum in the energy range 40-100 EeV, as measured by the Pierre Auger Observatory~\cite{PierreAuger:2020qqz,FENU20233531}. Specifically, as in Ref.~\cite{AlvesBatista:2019rhs}, the normalisation is chosen to match the measured EECR spectrum at $E_{0}=10^{19.55}$ eV. Since EECRs can only travel distances of the order of $\sim100$ Mpc before scattering, the measurement from Auger comes from very low redshift and thus the normalisation can be considered almost independent of the source evolution.

All in all, in our simplified modeling setup, the injected cosmogenic neutrino flux is given in terms of only three parameters: the spectral index of the injected proton spectrum ($\gamma$), the maximum value of the proton energy ($E_{\text{max}}$) and the source evolution parameter ($m$).

\begin{figure}[h!]
    \centering
    \includegraphics[width=0.9\linewidth]{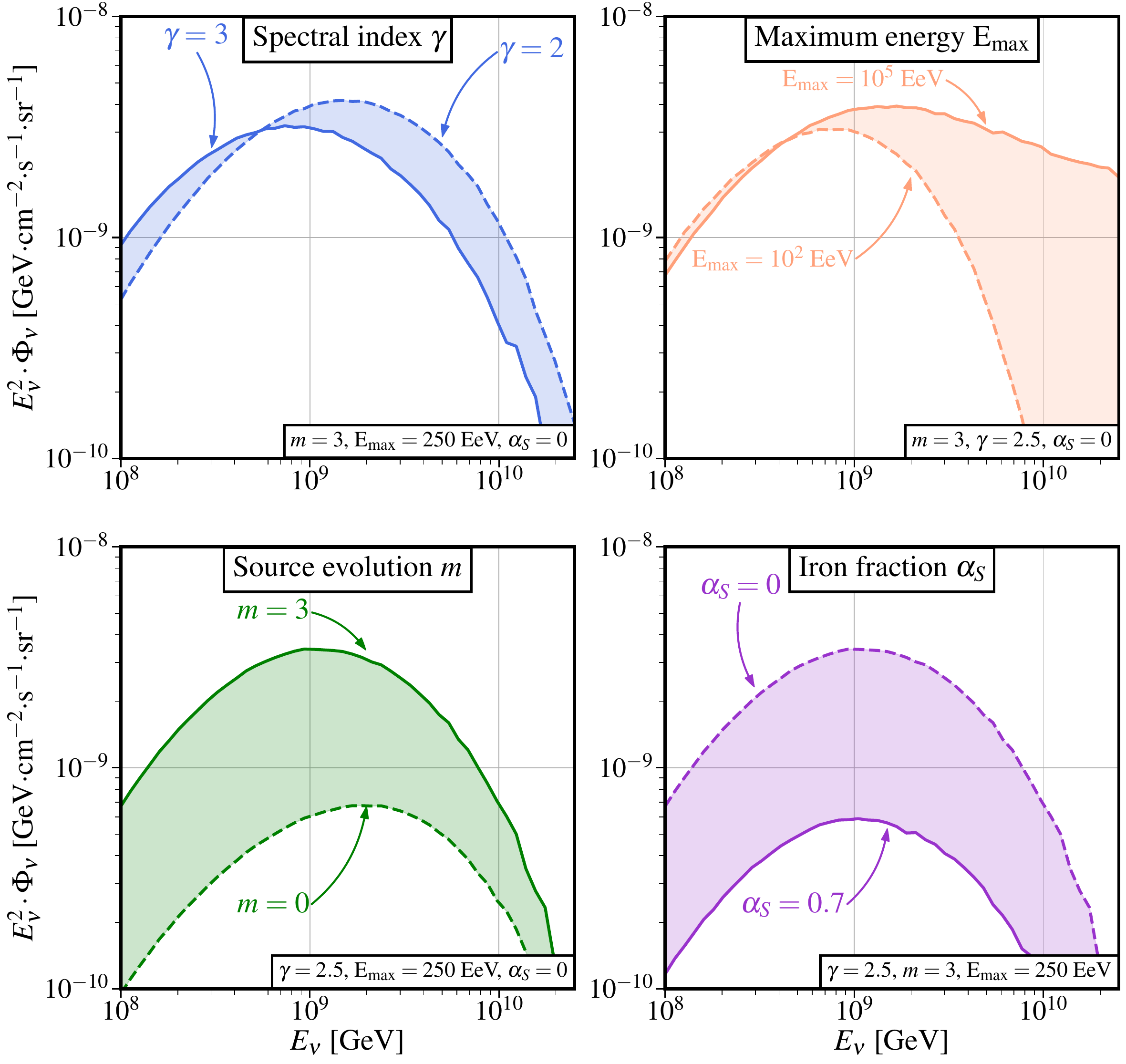}
    \caption{Simulated single-flavor cosmogenic neutrino spectra assuming the flavor ratio at the Earth to be $\nu_e:\nu_\mu:\nu_\tau=1:1:1$, for different values of the spectral index $\gamma$, maximum energy $E_{\text{max}}$, source evolution parameter $m$, and in the last panel, besides protons, we consider a fraction $\alpha_S$ of iron. Except for the last panel (bottom-right) all cases are for a pure-proton scenario.}
    \label{fig:varying-astrophysics}
\end{figure}

In Fig.~\ref{fig:varying-astrophysics}
we show the simulated  cosmogenic neutrino spectra~\footnote{The cosmogenic neutrino spectra we will present throughout this paper will always be 
the single-flavor one, i.e.,  $\nu_\tau$ and  $\bar \nu_\tau$ spectra combined.} at the Earth for  different values of the spectral index $\gamma =2$ and $3$ 
(top-left panel), of the maximum energy $E_{\text{max}}=10^{2}$ EeV and $10^{5}$ EeV (top-right panel) and  of the evolution parameter $m=0$ and 3 (bottom-left panel). We are assuming the flavor ratio at the Earth to be 
$\nu_e:\nu_\mu:\nu_\tau= 1:1:1$ in accordance with  what is expected from standard neutrino oscillation experimental results~\cite{SajjadAthar:2021prg} when the dominant neutrino production is $\pi$-decay. The parameters that are not varied are kept in their value of the benchmark point: $\gamma=2.5$,
$E_{\text{max}}=250$ EeV and $m=3$.
To illustrate the fact 
our parametrisation of the source evolution can mimic different chemical compositions, because of the degeneracy with $m$, we show in the bottom-right panel of Fig.~\ref{fig:varying-astrophysics} the simulated cosmogenic spectrum considering the source to produce, besides protons, a certain fraction $\alpha_S$ of iron. 

From this figure we can appreciate the effect of each astrophysical parameter on the spectrum, giving an indication of how our BSM results will depend on our assumptions on their  values. A lower (higher) spectral index shifts the peak of the spectrum to higher (lower) energies. Clearly, a lower (higher)  $E_{\text{max}}$ will produce a  lower (higher) cut-off, decreasing (increasing) the cosmogenic flux at higher energies. 
Note also that in the range  $2\times 10^8 \lesssim E_\nu/\text{GeV} \lesssim 2 \times 10^{9}$ the neutrino spectrum is almost not affected by the choice of $\gamma$ or $E_\text{max}$, this behavior does not depend on $m$.
The source evolution parameter (or the chemical composition) is the main responsible for the overall normalization, $m=3$ is a typical value for the  
star formation rate while $m=0$, flat evolution, is allowed by  some tidal disruption event and BL Lac models~\cite{vanVliet:2019nse}.

\section{Modeling of the Radio Array Detector Response}
\label{sec:detector}
We will consider the GRAND project~\cite{GRAND:2018iaj}, which 
aims to extend the field of radio-detection of EAS to include events initiated by Earth-skimming neutrinos, as the prototype for a radio array of antennas. This will  allow us 
to access the sensitivity of 
this type of detector to BSM physics that can affect 
the cosmogenic neutrino flux in a distinctive way.

In principle, at EeV energies the Earth is opaque even to Earth-skimming neutrinos. At those 
energies the neutrino-nucleon 
deep-inelastic scattering cross section 
increases as $E^{0.363}$~\cite{Gandhi:1998ri} so  the neutrino interaction length inside 
the Earth will be only of a few hundred kilometers. 
If the neutrino interacts via 
charged current, it will produce a  
charged lepton of the same flavor of 
the incoming neutrino. For $\nu_e$, 
the outgoing electron will give rise to
a short-lived electromagnetic shower
concentrated around  the point of 
interaction. Even if this shower starts 
close to the surface, the maximum column depth that the shower can traverse is about  30-40 m underground, making the expected effective volume for radio detection too small.
For $\nu_\mu$, the outgoing high energy $\mu$ can travel from the point of creation in the rock to the atmosphere
where it will have a range of several 
kilometers. The probability that this $\mu$ will 
decay above the radio array generating a detectable EAS seems also to be negligible. For $\nu_{\tau}$, however, 
things are different because of the properties of the $\tau$ lepton. First, it is the heaviest charged lepton so radiative losses are the most suppressed. Second, it has a short
lifetime ($\tau_\tau \sim 2.9 \times 10^{-13}$~s) which implies 
a range $\gamma \, c \, \tau_\tau \sim$ 50 km ($E_\tau/1$ EeV) before decaying.
Third, since the decay of a $\tau$ 
also produces a $\nu_\tau$, the attenuation of the neutrino flux due to interaction as it propagates in matter
is compensated by $\nu_\tau$ regeneration~\footnote{This may shift the flux to lower energies as the regenerated $\nu_\tau$ receives $\sim 30$\% of the parent $\tau$ energy.}.
So a $\tau$ produced by a $\nu_\tau$
underground can exit the Earth and decay in-flight in the atmosphere above the radio array
triggering a detectable EAS.  
As the EAS develops in the atmosphere, the geomagnetic field separates positive from negative charges  creating a time-dependent electric current which 
induces radio emission. Due to Compton scattering, an excess of negative charge builds up in the EAS during propagation
also inducing radio emission. This is know as the Askaryan effect~\cite{Askaryan:1962,Askaryan:1965}. 
In dense media, like ice, the Askaryan effect is more important, while in air the geomagnetic effect dominates.
Both emissions are coherent and beamed in the forward direction, inside a narrow cone of half-width equal to the Cherenkov angle.

Furthermore, most of the $\tau$ decays 
are into hadrons (65\%) or   
electrons (20\%) so once the $\tau$ reaches the atmosphere, the probability of creating a detectable air shower is rather high, i.e. only the $\tau$ decays into muons will not emit a significant radio signal (as in the case of $\nu_\mu$).

The idea is to instrument a large area using a sparse array. The strategy is to  divide  GRAND into 10 to 20 geographically  separate and favorable to neutrino detection independent sub-arrays, each containing about $10^{4}$ antennas. Ideally, the antenna array should be deployed on a mountain slope facing another mountain that could serve 
as an additional target for interaction. This mountain should be distant enough (few tens of km) for the EAS to develop and the radio cone to become large enough before hitting the antennas. 

GRAND is expected to have a very good (sub-degree) angular resolution.
This can, among other things, help to reject background events coming from above the horizon. Preliminary estimates indicate that one expects less than 0.1 per year ultra-high-energy cosmic ray mis-identified as a neutrino  in GRAND200k, a combined area of   200 000 km$^2$, with the angular aperture $85^\circ\leq\theta\leq 95^\circ$~\cite{GRAND:2018iaj}.

It should also be mentioned that not all cosmogenic neutrino observatories will have the same detection techniques. For instance, the IceCube-Gen2 radio array~\cite{IceCube-Gen2:2020qha,IceCube-Gen2:2021rkf} will be embedded in the ice; therefore, it will be sensitive to interactions of cosmogenic neutrinos of any flavor. In fact, the different flavor sensitivity of these observatories may be harnessed to constrain the production mechanisms of cosmogenic neutrinos~\cite{Testagrossa:2023ukh}.

\begin{figure}[h!]
    \centering
    \includegraphics[width=0.9\linewidth]{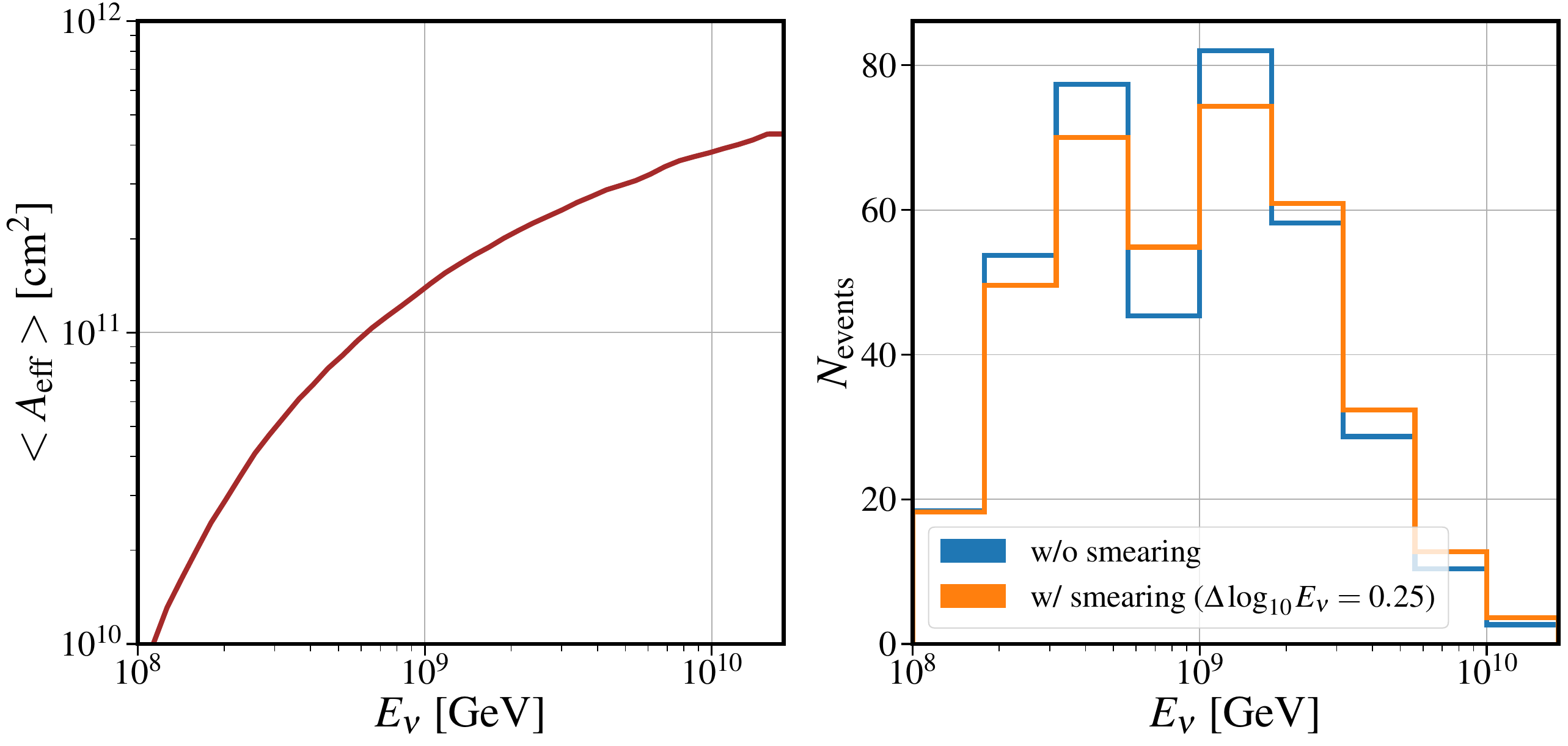}
    \caption{Direction-averaged effective area (left panel) and effect of the energy resolution in the distribution of events (right panel). The effective area has been taken from Ref.~\cite{GRAND:2018iaj}, while for the energy resolution we demonstrate its effect by means of a mock flux in which a spectral feature (a dip) has been artificially added so as to mimic the possible signatures of the BSM models under consideration.}
    \label{fig:effarea_and_resolution}
\end{figure}

In order to convert the calculated cosmogenic neutrino energy spectrum into 
the number of $\nu_\tau$ events 
detected at GRAND, one needs to know the effective area of the array. The GRAND collaboration has already simulated the effective area of {\em HotSpot 1} (HS1), 
a 10 000  km$^2$ area located at the Southern rim of the Tian Shan mountain range in China, thus having an exceptionally good exposure to neutrinos. We will 
make use of the direction-averaged effective area shown in Fig. 25 of~\cite{GRAND:2018iaj} as a GRAND200k~\footnote{The ultimate goal of GRAND is to have 20 similar setups.} estimation, obtained by multiplying by 20 the HS1 effective area with an aggressive energy threshold for the radio antennas~\footnote{In this case the antennas are triggered if the peak-to-peak amplitude of the voltage signal  exceeds 30 $\mu$V, instead of the conservative 
threshold of 75 $\mu$V, increasing the effective area by a factor of 2.5 with respect to the conservative case. 
This also gives twice the expected background noise in the 50-200 MHz frequency range of the antennas.}. On the left panel of Fig.~\ref{fig:effarea_and_resolution} we show the effective area we use in this work.

The energy resolution of the reconstructed EAS is expected to be of 15\%. However, 
it is initiated by a $\tau$ which carries and unknown fraction of the energy of the parent $\nu_\tau$.
So to estimate the neutrino energy resolution, one first needs to reconstruct the energy of the shower and then unfold into the neutrino energy, taking into account the fraction of energy carried by the $\tau$ as well as
the energy losses of $\nu_\tau$ and $\tau$ in matter. It seems plausible to 
assume an  neutrino energy resolution of $\Delta \log_{10}E_{\nu}=0.25$, as proposed in~\cite{Moller:2018isk}.
In our simplified detector simulation, we will estimate the number of 
$\nu_\tau$ events at GRAND by 
smearing the  cosmogenic 
$\nu_\tau$ spectrum, calculated according to each BSM scenario (see section \ref{sec:BSM}), times the  effective area with a log-normal distribution adopting this energy resolution as the width. We will collect the number of events in 9 log-energy bins of width $\Delta \log_{10}E_{\nu}=0.25$, from $E_\nu=10^{17}$ eV to $E_\nu=10^{19.25}$ eV. On the right panel of Fig.~\ref{fig:effarea_and_resolution} we illustrate the effect of the energy resolution on the distribution of cosmogenic neutrino events. In particular, we have artificially added a dip of width $\Delta \log_{10} E_{\nu}=0.25$ at $E_\nu=0.7$ EeV to mock a spectrum which mimics the typical signature of some of the BSM scenarios that will be studied. The plot highlights the relevance of including the energy resolution, as it tends to smooth out any spectral feature, thus significantly reducing the sensitivity.

\section{BSM Scenarios}

We have identified a few interesting BSM scenarios which can   modify the expected cosmogenic $\nu_\tau$ energy spectrum at GRAND in a distinct observable way, which cannot be imitated by a simple renormalization of the distribution.  

\label{sec:BSM}
\subsection{Neutrino self-interactions ($\nu$SI) }
\label{subsec:self}

Secret neutrino-neutrino interactions~\cite{Beacom:2004yd} 
can significantly alter the propagation of cosmogenic neutrinos, as they can induce scattering with relic neutrinos from the Cosmic Neutrino Background ($\cnb$) radiation leaving visible imprints on the cosmogenic $\nu_\tau$ spectrum~\cite{Fiorillo:2020jvy}, as the baselines traveled by these neutrinos are very large. We focus on the scalar interaction case, as it is the least constrained setup today. 
As such, we consider the following Lagrangian to describe these 
interactions with Majorana neutrinos~\footnote{We use this effective Lagrangian description which is enough for our study.
Possible gauge invariant UV complete models that can give rise to this can 
be found, for instance, in  \cite{Blum:2014ewa,Berryman:2018ogk,Kelly:2020pcy}. Note that $\phi$ can only couple to the SM fields via 
dimension six or higher   operators~\cite{Berryman:2022hds}.}

\begin{equation}
    -\mathcal{L}\supset \dfrac{1}{2}g_\phi^{\alpha\beta} \, \overline{\nu_{\alpha L}} \, \nu_{\beta L}^{c} \phi+{\rm h.c.}+\dfrac{1}{2}m_{\phi}^2\phi^2\, ,
\end{equation}
where $\phi$ is the real scalar mediator with mass $m_\phi$ and we denote the flavor indices as $\alpha,\beta = e,\mu,\tau$. Note that, at the energies 
we are interested in, scalar and 
pseudoscalar interactions give the same results. This scalar mediator is a SM singlet and carries a +2 $(B-L)$ charge.
Since relic neutrinos are non-relativistic, the center-of-mass energy squared of the scattering of a cosmogenic neutrino of energy $E_\nu$ and a relic neutrino will be given by $s=2E_\nu m_i$, where $m_i \lesssim 0.1$ eV is the
neutrino mass. When the condition
\begin{equation}
    s=2E_\nu m_i=m_\phi^2\, ,
\end{equation}
is met, the cross section is resonantly enhanced and can lead to a dip in the spectrum of cosmogenic neutrinos with  energies satisfying the above condition, 
as the Universe becomes opaque to those 
neutrinos for large enough couplings~\cite{Esteban:2021tub}. We will always assume that all neutrino mass eigenvalues are larger than the temperature of the C$\nu$B so that all cosmic neutrinos can be considered non-relativistic in the redshift interval of interest. For the sake of simplicity, we will assume  Normal Ordering and a value of $\sum m_i=0.1$ eV, although results for a different $\sum m_i$ will be qualitatively similar, as long as all $m_{\rm lightest}\gg T_{\text{C}\nu \text{B}}$. It should be noted that, if the lightest neutrino mass eigenstate lies below $T_{\rm C\nu B}\simeq 1.7\cdot 10^{-4}$ eV, its Fermi-Dirac distribution would allow for a broadening of the neutrino absorption feature, thus increasing the sensitivity to $\nu$SI~\cite{Wang:2025qap}.

Therefore, provided that GRAND is able to measure a sufficient number of $\nu_{\tau}$ events in order to reconstruct their energy spectrum, it should have sensitivity for mediators whose masses can satisfy the resonance condition, i.e. 
100 MeV $\lesssim m_\phi \lesssim 1$ 
GeV, for $E_\nu$ in the range 0.1 to 10 EeV.

For the sake of simplicity, we will restrict ourselves to a mediator that couples predominantly to tau neutrinos, i.e. $g_\phi^{\alpha\beta}=g_{\tau\tau}\, \delta_{\alpha\tau}\delta_{\beta\tau}$. 
This approach is also justified 
because $\nu_\tau$ self-interactions are the least explored as the scalar coupling to electron and muon neutrinos are significantly more 
constrained in the $m_\phi$ range we are exploring here. 
Furthermore, the authors of Ref.~\cite{Blinov:2019gcj}
 pointed out that neutrino self interaction can explain the tension 
in the Hubble parameter if $m_\phi \sim$
10 MeV and has large couplings (almost)
exclusively to $\nu_\tau$. 
In this simplified setup, our $\nu$SI model is described by only two parameters: the mediator mass, $m_\phi$, and its coupling to $\nu_\tau$, $g_{\tau \tau}$.

On their way  to Earth,  cosmogenic $\nu_\tau$ and $\bar \nu_\tau$ may scatter with the C$\nu$B due to $\nu$SI. 
As a consequence, higher-energy neutrinos are absorbed and lower-energy neutrinos are regenerated. 
One can estimate the size of the cross section $\sigma_{\nu \rm SI}$ 
which can significantly affect propagation by requiring that the optical depth for this interaction as 

$$\uptau_\nu \simeq \frac{n_{\rm C\nu B}\;  \sigma_{\nu \rm SI}}{h_0} = 1\, ,$$
where $h_0=  67.3$ km s$^{-1}$ Mpc$^{-1}$ 
is the Hubble parameter (we 
consider $h_0^{-1}$ as the typical traveled distance scale from the astrophysical source~\footnote{We are using natural units throughout this paper, so $h_0^{-1}$ is in fact $c/h_0$, i.e. $h_0^{-1} \sim 4.3$ Gpc ($ \sim 10^{23}$ km).}) and $n_{\rm C\nu B} \sim 56$ cm$^{-3}$~\cite{DiValentino:2024xsv} is the target C$\nu$B number density for $\nu_\tau$ (or $\bar \nu_\tau$). This results in a cross section $\sigma_{\nu \rm SI} \sim 10^{-30}$ cm$^{2}$, much  
higher than  $\sigma_{\nu \rm SM} \simeq G_F^2 s \sim 10^{-37} \times (E_\nu/{\rm EeV})\times (m_\nu/{\rm eV})$ cm$^2$  expected for SM neutrino-neutrino scattering.

In order to estimate how $\nu$SI modifies the cosmogenic neutrino flux, we will make use of the code \texttt{nuSIprop}~\cite{Esteban:2021tub}, which solves the transport equation of high-energy neutrinos (see Eq.~\eqref{eq:transp}) in the presence of self interactions. In this case 
the interaction terms to be included in the transport equation~\cite{Ng:2014pca,Creque-Sarbinowski:2020qhz} are
\begin{align}
    \Gamma_\nu (z,E(z,E_\nu)) =  n_{\rm C\nu B}(z) \, \sigma_{\nu \rm SI}(E(z,E_\nu))\, ,
    \end{align}
and 
\begin{align}
      \int dE' \; \frac{\sigma_{ji}}{dE} \to  n_{\rm C\nu B}(z) \int_{0}^{\infty} dE_\nu' \,   \frac{d\sigma_{\nu \rm SI}}{dE_\nu} (E_\nu,E_\nu')\, ,
     \end{align}
where $\sigma_{\nu \rm SI}$ is the $\nu \nu \to \nu \nu$ scattering cross-section that can be found in Appendix A of Ref.~\cite{Esteban:2021tub}.

Since \texttt{nuSIprop} considers a power-law injection spectrum for the neutrinos, which is not valid for cosmogenic neutrinos, we have adapted the code to take into account the cosmogenic spectral shape and its redshift dependence. For this purpose, we simulate the cosmogenic neutrino spectrum with CRPropa3 as described in the previous section, and then we interpolate its shape both in energy and redshift. This interpolated spectrum is then fed into \texttt{nuSIprop}, which propagates the neutrinos taking $\nu$SI into account.

\begin{figure}[h!]
    \centering
    \includegraphics[width=0.9\linewidth]{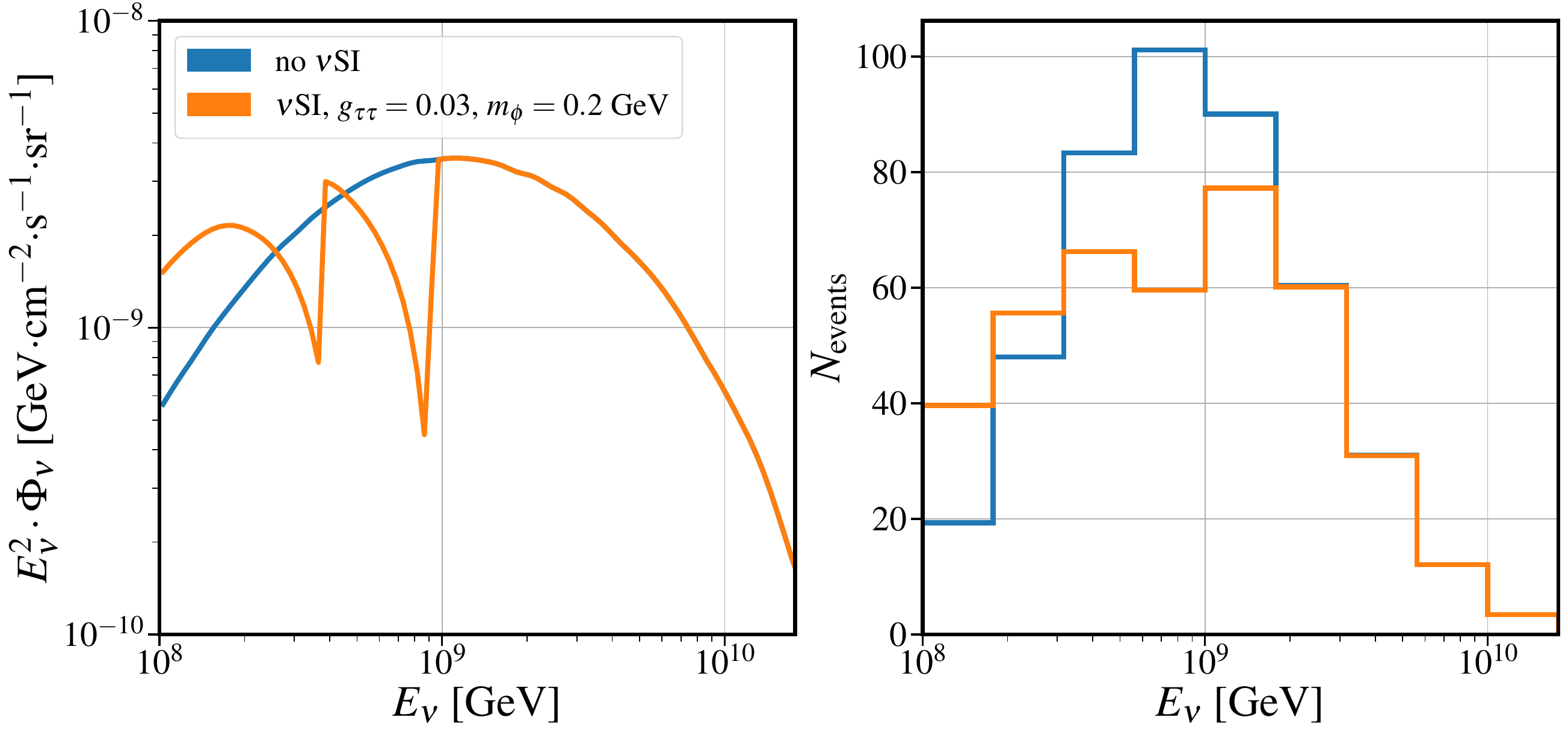}
    \caption{On the left panel we present the single-flavor cosmogenic neutrino flux at the Earth for $\gamma=2.5,    m=3$ and $E_{\rm max }=250$ EeV without $\nu$SI (blue) and with $\nu$SI for $g_{\tau\tau}=0.03$ and $m_\phi =0.2$ GeV (orange). On the right panel we have the corresponding event distributions at GRAND (10-year exposure).}
    \label{fig:example_flux_nuSI}
\end{figure}

For a given point of the parameter space
$(m_\phi,g_{\tau \tau},\gamma,m,E_{\rm max})$ we compute
the modified cosmogenic neutrino 
flux arriving at the Earth. In Fig.~\ref{fig:example_flux_nuSI} we illustrate the effect of $\nu$SI in the neutrino flux (left panel) as well as the observed number of events (right panel). The double dip feature shown in the right panel corresponds to the energies in which the resonance condition $2E_{\nu}m_i=m_\phi^2$ is met. Since we have assumed Normal Ordering, the dip at lower energies corresponds to scatterings with $\nu_3$, which is the heavier mass eigenstate, while the dip at higher energies corresponds to the lighter $\nu_1$ and $\nu_2$. This also explains why the higher energy dip is deeper, as more neutrino states can satisfy the corresponding resonance condition. Additionally, since the resonance condition can be met at different redshifts, the dip is broadened to smaller energies. Lastly, notice the bump at energies below the dips, which is due to energy conservation: for every interaction, a cosmogenic neutrino  gets downscattered while a C$\nu$B neutrino gets upscattered. All in all, the typical spectral features arising from $\nu$SI are a depletion of the flux for some energies followed by an enhancement at lower energies. As pointed out in Ref.~\cite{Esteban:2021tub}, this double-dip plus bump spectral feature is difficult to explain away with modified astrophysics or random fluctuations since the position of the dips and the bump are mostly fixed by external data i.e. neutrino oscillations and cosmology.

\subsection{Pseudo-Dirac neutrinos (PD$\nu$)}
\label{subsec:pseudo-dirac}

Another BSM scenario that could potentially leave an observable imprint in the cosmogenic neutrino spectrum is if neutrinos have a pseudo-Dirac nature~\cite{Wolfenstein:1981kw,Petcov:1982ya,Valle:1983dk}. 
Pseudo-Dirac fermions are fundamentally Majorana fermions which behave basically like Dirac particles for terrestrial experiments because of the extremely small mass-squared difference $\delta m^2$ between their active and sterile components.
Theoretically, this scenario amounts to neutrinos receiving most of their mass from a Dirac mass term $m_D$, but also having a small Majorana mass $\mu$, the mass Lagrangian being 
\begin{equation}
    -\mathcal{L}\supset m_D \overline{\nu}_L\nu_R+\mu \, \overline{\nu_R^{c}}\,\nu_R\, ,
\end{equation}
where $\nu_L$ are left-handed neutrinos, part of the SM lepton doublets, and $\nu_R$ are right-handed SM singlets.
The small $\mu$ lifts the degeneracy of the mass eigenvalues, so in the end we have pairs of eigenstates with almost degenerate masses. Since this also
entails a soft breaking  of the accidental SM lepton number symmetry, one could argue that 
pseudo-Dirac  neutrinos 
could be connected to quantum gravitational effects, which are expected to break  global symmetries~\cite{Kallosh:1995hi}. Furthermore, there are many well-motivated models in the literature that can predict pseudo-Dirac neutrinos~\cite{Chang:1999pb,Joshipura:2000ts,Lindner:2001hr,Berezinsky:2002fa,Ma:2014qra,Valle:2016kyz,Ahn:2016hhq,Babu:2022ikf}.

In this scenario, active-sterile oscillations are expected to modify the standard oscillations~\cite{Kobayashi:2000md} leading to observable effects in experiments~\cite{Martinez-Soler:2021unz,Franklin:2023diy,Carloni:2022cqz,
Dev:2024yrg}. The effect is proportional to the tiny mass squared splitting $\delta m^2$ so one needs astrophysical to cosmological distances for these oscillations to develop. This makes cosmogenic neutrinos interesting to search for these oscillations.

The evolution of cosmogenic neutrinos in vacuum is governed by the Hamiltonian:
\begin{align*}
    H_{\rm vac}(z) = \frac{1}{E_\nu(z,E_\nu)}V^\dagger M^2_{\rm diag} V\, ,
\end{align*}
where $E_\nu(z,E_\nu)=E_\nu(1+z)$ is the neutrino energy at redshift $z$ when $E_\nu$ is the neutrino energy at 
the Earth and 
$M^2_{\rm diag}$ is a $6\times 6$
 diagonal mass matrix with entries  
\begin{align}
    m^2_{i \pm} = m_i^2 \pm \frac{\delta m_i^2}{2}\, , \quad i=1,2,3,
\end{align}
which correspond to the mass eigenstates which are a maximal mixture of active and sterile components 
\begin{eqnarray}
    \nu_{i}^+ &=& \frac{1}{\sqrt{2}} (\nu_{ia} + \nu_{is})\, , \\
     \nu_{i}^- &=& \frac{-i}{\sqrt{2}} (\nu_{ia} - \nu_{is})\, ,
     \end{eqnarray}
so the $6 \times 6$ unitary matrix $V$ can be parametrized as~\cite{Kobayashi:2000md} 
\begin{align}
    V = \frac{1}{\sqrt{2}}\left( \begin{array}{cc} U & 0 \\ 0  &U_{\rm N}\end{array} \right) \left(\begin{array}{cc}\mathbf{I_3} & i \mathbf{I_3} \\ \phi & -i  \phi \end{array} \right)\, ,  
\end{align}
  where $U$ and $U_{N}$ are, respectively, the PMNS  and another unitary mixing matrix that diagonalize the active and sterile sectors, $\phi =\rm{diag}(e^{-i\phi_1}, e^{-i\phi_2}, e^{-i \phi_3})$ is a diagonal 
phase matrix and $\mathbf{I}_3$ is the $3\times 3$ identity matrix. The flavor 
states  can be written in terms of the
almost degenerate mass eigenstates as 
\begin{align}
    \nu_\alpha = \frac{U_{\alpha i}}{\sqrt{2}} (\nu_i^+ + i \, \nu_i^-), \quad \alpha = e,\mu,\tau
\end{align}
so the flavor oscillation probabilities 
will, in general, depend on the two larger mass splittings $\Delta m^2_{21} \equiv m^2_2-m^2_1 \simeq m^2_{2\pm}-m^2_{1\pm} \sim 7.5 \times 10^{-5}$ eV$^2$ and 
$\vert \Delta m^2_{31} \vert \equiv \vert m^2_3 -m^2_1\vert \simeq \vert m^2_{3\pm} - m^2_{1\pm}\vert \sim 2.5 \times 10^{-3}$ eV$^{2}$~\cite{Esteban:2024eli} as well as on the active-sterile splitting $\delta m_i^2$. Since $\delta m_i^2 \ll \Delta m^2_{21},\vert \Delta m^2_{31}\vert$ we will assume, for simplicity, they are all degenerate and equal to $\delta m^2$.
Furthermore, the oscillation length is defined as 
\begin{align}
    L_{\rm osc} = \frac{4 \pi E_\nu}{\Delta m^2} \approx 8 \times 10^{-9} \,  \left( \frac{E_\nu}{1 \, \rm EeV}\right) \left(\frac{10^{-5}\, \rm eV^2}{\Delta m^2} \right) \, {\rm Gpc}\, ,  
\end{align}
where $\Delta m^2= \Delta m^2_{21}, \vert \Delta m^2_{31}\vert$ or $\delta m^2$. This is much smaller than a few Gpc (the typical scale distance traveled by a cosmogenic neutrino to Earth) for the two larger mass squared differences, so these oscillations are averaged out over cosmological distances.    
This also tells us that cosmogenic neutrino data can probe  $\delta m^2 \sim 10^{-14}$ eV$^2$.
Also, because the effective distance due to redshift traveled by cosmogenic neutrinos from their production point (at $z_p$) to  Earth (at $z=0$)  can be computed as  
\begin{align}
    L_{\rm eff} (z_p)= \int_{0}^{z_p} \frac{dz}{H(z)(1+z)^2} \, ,
    \label{eq:effective_length}
\end{align}
the  final $\nu_\alpha$ to $\nu_\beta$ oscillation probability will 
be simply
\begin{align}
    P(\nu_\alpha \to \nu_\beta; z_p) = \frac{1}{2} \sum_{i=1,3} \vert U_{\alpha i}\vert^2 \vert U_{\beta i}\vert^2 \left[ 1+ \cos\left(\frac{\delta m^2 L_{\rm eff}(z_p)}{2 E_\nu}\right)\right]\,,
    \label{eq:probability_PDnu}
\end{align}
and similarly for the antineutrino modes.
This will impact the cosmogenic $\nu_\tau+\bar \nu_\tau$ flux that can be measured by GRAND. For the PMNS entries $U_{\alpha i}$, we fix the mixing angles and phase to the latest bestfit values from the global neutrino oscillation analysis~\cite{Esteban:2024eli}.

It should be noted that, as pointed out in Ref.~\cite{Dev:2024yrg}, pseudo-Dirac neutrinos traveling through the Universe feel a matter effect with the C$\nu$B due to coherent $Z$-exchange, as the right-handed component is sterile under weak interactions. In the interaction basis, this matter Hamiltonian reads:
\begin{equation}
    H_{\rm mat}(z) = \frac{G_F}{\sqrt{2}} n_{\text{C}\nu\text{B}}(z)
\left(\begin{array}{cc}\mathbf{I_3} & 0 \\ 0 & 0 \end{array} \right)\,,
\end{equation}
where $n_{\text{C}\nu\text{B}}(z)=n_{\text{C}\nu\text{B}}^{(0)}(1+z)^3$ C$\nu$B number density at redshift $z$, which reduces as the universe expands.
As stated in Ref.~\cite{Dev:2024yrg}, this matter effect tends to drive the active-sterile mixing out of the maximal regime, thus damping the oscillatory pattern that could be resolved in neutrino telescopes. However, the size of this effect is negligible when compared with the vacuum term if no sizable C$\nu$B overdensities are considered. Indeed, the relative size of the matter and vacuum Hamiltonian is given by:

\begin{equation}
    \frac{||H_{\rm mat}||}{||H_{\rm vac}||}\sim 1.4\times 10^{-3} \left(\frac{10^{-14}\, {\rm eV^2}}{\delta m^2}\right)\left(\frac{E_\nu}{1\,{\rm EeV}}\right)\left(1+z\right)^4\, .
\end{equation}

Notice how the different redshift dependence of the vacuum 
and matter Hamiltonian makes the matter effect stronger 
at higher redshifts. In order to make sure that neglecting the
matter effect is a good approximation in all of the parameter
space relevant for GRAND, we have performed the full computation
of the oscillation probability, that is, we have explicitly solved the time-dependent Schrödinger equation
$\frac{d}{dt}\vert\nu_a(t)\rangle
=-i [H_{\rm vac}(t)+
H_{\rm mat}(t)]_{ab}\vert\nu_b(t)\rangle$,
finding that $H_{\rm mat}$ can be safely neglected in the 
$\delta m^2$ interval relevant for our analysis.

\begin{figure}[h!]
    \centering
    \includegraphics[width=0.9\textwidth]{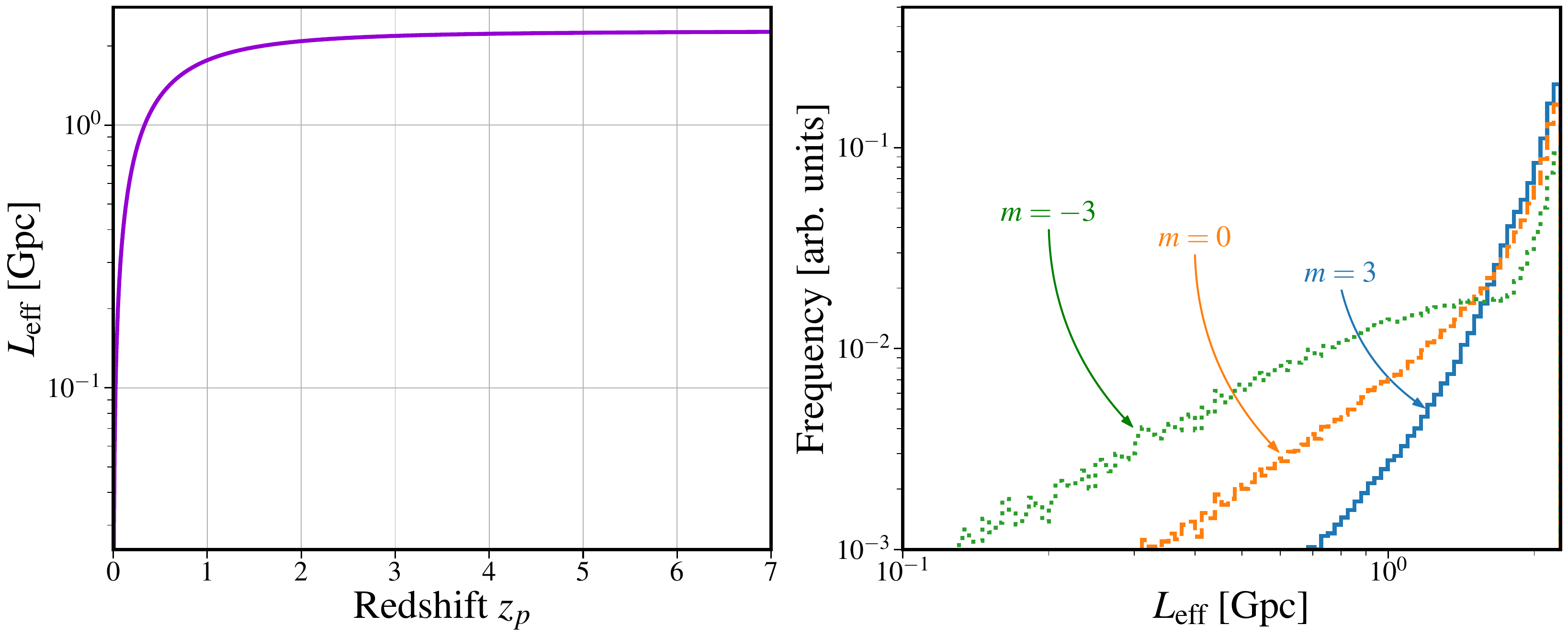}
    \caption{On the left panel we show how $L_{\rm eff}$ changes as a function of the redshift $z_p$ of the production point. On the right panel we can see for different values of the source evolution parameter $m=-3$ (dotted green), $m=0$ (dashed orange) and $m=3$ (solid blue) the distribution of events as a function of $L_{\rm eff}$, where we fixed $\gamma = 2.5$ and $E_{\rm max} = 250$ EeV.}
    \label{fig:effective_length}
\end{figure}

Let us  now discuss  the robustness of the bounds that can be derived on the PD$\nu$ mass-splittings $\delta m^2$ from the diffuse analysis. Indeed, the optimal method of deriving robust bounds on $\delta m^2$ would be to identify specific point sources at known redshifts, so that the effective baseline~\eqref{eq:effective_length} is fixed and the only free parameter in the survival probability is $\delta m^2$. However, when analyzing the diffuse flux, assumptions on the source distribution must be taken, which might render the derived constraints on $\delta m^2$ dependent on said assumptions. However, here we argue that this model-dependence is not as pronounced as it would naively seem. The main reason why this is the case can be seen from the left panel of Fig.~\ref{fig:effective_length}, where we plot the effective neutrino baseline $L_{\rm eff}$ as a function of the redshift $z_p$. Indeed, $L_{\rm eff}$ quickly saturates to $\sim 2$ Gpc such that all neutrinos coming from $z_p\gtrsim0.5$ effectively travel the same distance from the point of view of the oscillation probability. Alternatively, in the right panel, we show the (normalized) $L_{\rm eff}$ distribution of simulated events for three different source evolutions (see Eq.~\eqref{eq:source_evolution}): $m=3$ (solid blue), $m=0$ (dashed orange) and $m=-3$ (dotted green). In all of the three cases, the distributions are clearly peaked at $L_{\rm eff}\simeq 2$ Gpc, with the sharpness of the peak decreasing with $m$, which can be intuitively understood as a smaller $m$ increasingly suppresses sources at redshifts for which $L_{\rm eff}\sim$ Gpc. However, even for the weakest source evolution considered in Fig.~\ref{fig:effective_length}, i.e. $m=-3$, the fraction of neutrinos whose $L_{\rm eff}\in[1,2.25]$ Gpc is $65\%$, which increases to $86\%$ and $97\%$ for $m=0$ and $m=3$, respectively. Therefore, for a wide range of source evolutions, a large proportion of neutrinos have very similar $\sim$ Gpc oscillation baselines, making the analysis of PD$\nu$ quite robust even in the context of a diffuse flux measurement.

\begin{figure}[h!]
    \centering
    \includegraphics[width=0.9\linewidth]{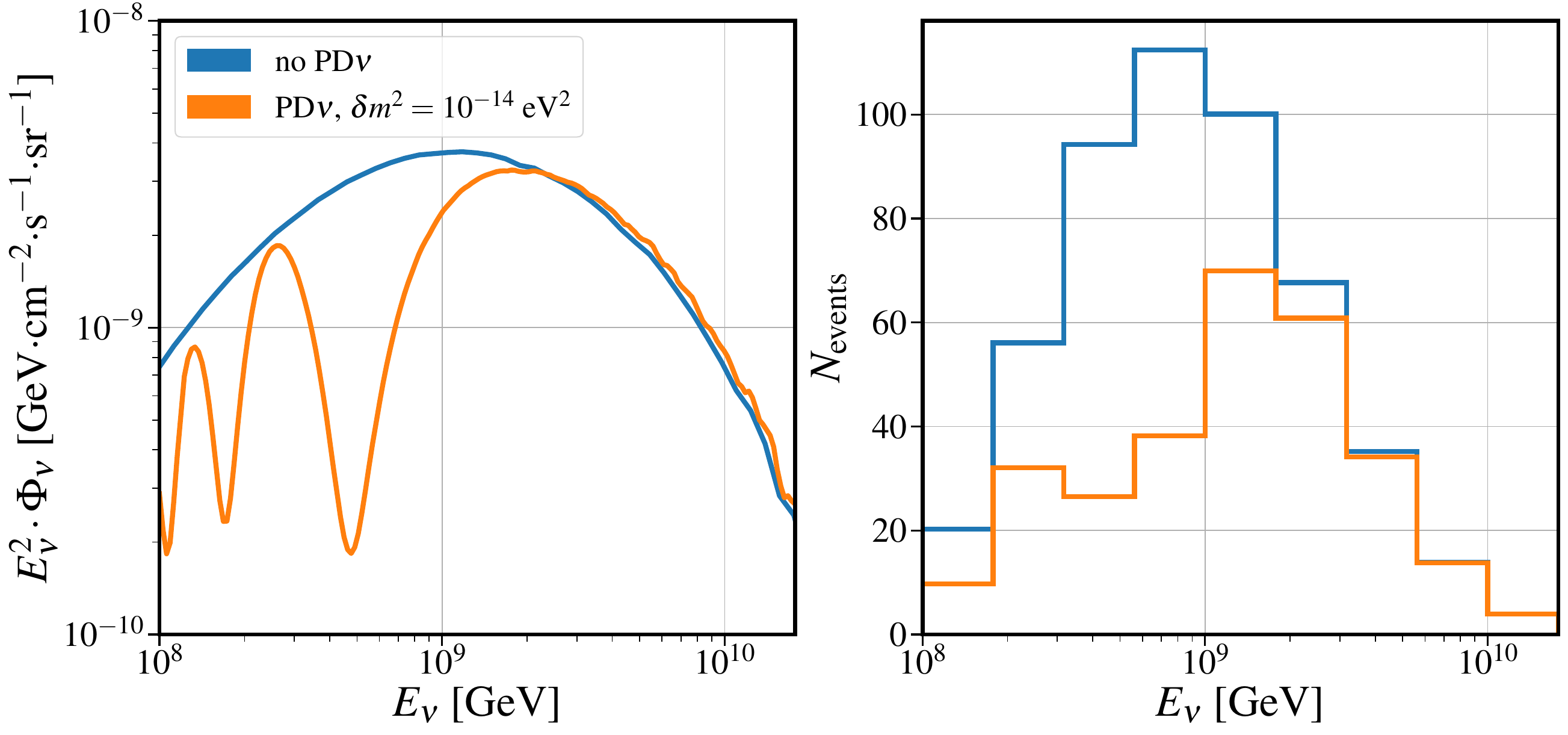}
    \caption{On the left panel we present the single-flavor cosmogenic neutrino flux at the Earth for $\gamma =2.5, m=3$ and $E_{\rm max}= 250$ EeV without PD$\nu$ (blue) and with PD$\nu$ for $\delta m^2 = 10^{-14}$ eV$^{2}$ (orange). On the right panel we have the corresponding event distributions at GRAND (10-year exposure).}
    \label{fig:example_pseudodirac}
\end{figure}

We compute the cosmogenic flux in the presence of PD$\nu$ by simulating the production of cosmogenic neutrinos as outlined in Sec.~\ref{sec:flux} and reweighting each event by its corresponding survival probability, given by Eq.~\eqref{eq:probability_PDnu}. In Fig.~\ref{fig:example_pseudodirac} we compare the cosmogenic flux and expected number of events at GRAND considering SM physics (blue) and PD$\nu$ (orange). It is clear that active-sterile oscillations produce an oscillatory pattern of dips in the spectrum when the condition $\delta m^2 L_{\rm eff}/(2 E_\nu)\simeq (2n+1)\pi$ is met, since the active-sterile conversion is maximal. The first oscillation dip (the one appearing at higher energies) is of special relevance in order to disentangle the effect of PD$\nu$ from astrophysics or statistical fluctuations. Indeed, it is the widest dip and thus the easiest to resolve, given the energy resolution. As it can be seen from Fig.~\ref{fig:example_pseudodirac}, only the first oscillation dip is appreciable in the distribution of the number of events, while subsequent oscillations are not resolved and instead amount to an overall decrease on the number of events at the lowest energy bins. In contrast to the $\nu$SI scenario, the spectral dip generated by PD$\nu$ is not accompanied by a bump and its position is not given by any external input, which makes the sensitivity very dependent on observing a suppressed number of events in some energy bins. This kind of signal, while not necessarily easy to mimic by varying the astrophysics, does become susceptible to random underfluctuations if the flux measurement is statistically limited.

\subsection{Neutrinos Scattering on Ultra-light Scalar Dark Matter ($\nu$DM)}
\label{subsec:nuDM}
There is abundant evidence for the existence of Dark Matter (DM) coming from
observations related to gravity that span a wide range of scale distances, from galaxy sizes  to the observable Universe. Yet we have no clue about the particle nature of DM and how the dark sector communicates with the SM one. An interesting possibility 
is that DM could be made of ultra-light scalars, a conceivable  fuzzy DM candidate. 
Being macroscopically delocalized,
fuzzy DM can potentially alleviate 
some problems of $\Lambda$CDM related 
to small-scale: the cusp vs. core, the missing satellites, and the too-big-to-fail problems~\cite{Brdar:2017kbt}. The production of such a light DM could be achieved in the early Universe, for instance, by means of the misalignment mechanism~\cite{Abbott:1982af,Dine:1982ah,Preskill:1982cy}.

If neutrinos  provide a portal to these ultra-light scalar fields, a general effect is that their masses and mixings  receive corrections that are time-dependent, potentially affecting  neutrino oscillations~\cite{Chun:2021ief,Berlin:2016woy,Krnjaic:2017zlz,Brdar:2017kbt,Capozzi:2018bps,Dev:2020kgz,Losada:2021bxx,Dev:2022bae,Davoudiasl:2023uiq,Gherghetta:2023myo,Sen:2023uga}. Furthermore, it has also been proposed that neutrino interactions with ultra-light dark matter could be responsible for the phenomenon of neutrino oscillations even in the presence of massless neutrinos~\cite{Sen:2023uga}. In this kind of scenario, neutrinos would acquire a ``refractive'' mass from the matter effect with the DM, leading to very rich phenomenology, as neutrino masses would not only showcase an energy dependence but also depend on the dark matter density. One of the best probes for such interactions is the measurement of astrophysical neutrinos propagating through cosmological distances, as they traverse large column depths of DM on their way to Earth. Since this kind of interaction could render the Universe opaque to these neutrinos, the observations of Supernova neutrinos~\cite{Kamiokande-II:1987idp,Hirata:1988ad,Bionta:1987qt,Alekseev:1988gp}, as well as high-energy neutrinos at IceCube~\cite{IceCube:2018dnn} can be employed to place constraints on the parameter space~\cite{Esteban:2021tub}. Alternatively, interactions of cosmic neutrinos with DM during the evolution of the Universe can leave an imprint on the CMB and matter power spectrum, which can be harnessed to constrain $\nu$DM interactions~~\cite{Bertoni:2014mva,Wilkinson:2014ksa,Escudero:2018thh,Brax:2023tvn}.

We will consider here a model where neutrinos  interact  with  ultra-light scalar bosons $S$ and their antiparticles $\bar S$, which could act as cold (non-relativistic) DM, via a new fermion $\chi$ that could be either a Dirac or a Majorana particle~\cite{Sen:2023uga}.  
A single scalar is enough for our purposes here as we do not require the ultra-light scalars to explain neutrino masses and mixings.
The Lagrangian describing neutrino scattering on DM in the mass basis can take the form~\footnote{This kind of effective Lagrangian can be accomplished 
in a gauge invariant way by UV-complete models 
which may involve, for e.g.,  pairs of vector-like fermions which are SM singlets or SM triplet  fermions (see Ref.~\cite{Blinov:2019gcj} and references therein).} 
\begin{align}
{\cal L} \supset    - y_{k} \, \bar \chi_R  \, \nu_{k L} \, S + y^*_{k} \, \bar \nu_{k L} \, \chi_R \, S^* \, ,
    \label{eq:ScalarDM}
\end{align}
with $k=1,2,3$ . The DM particles have mass $m_S \ll$ eV and 
the fermions $\chi$ have mass $m_\chi \gg m_S$. The scalars' contribution to the   energy density of the Universe is
\begin{align}
    \rho_S = m_S (n_S + n_{\bar S})\,, 
\end{align}
where $n_S$ ($n_{\bar S}$) is the scalar $S$ ($\bar S$) number density, and we will suppose $\rho_S$ to be equal to the entire cosmological DM energy density $\Omega_{\rm DM}=0.25$ when solving the extragalactic transport equation (see Appendix~\ref{app:transport}). It should be noted that we will only consider the interactions of cosmogenic neutrinos with the cosmological dark matter abundance. Even though galactic scales are much smaller than cosmological scales, the high DM overdensity inside the Milky Way's halo compensates for this hierarchy, thus rendering the galactic and cosmological DM column depths comparable. For the sake of simplicity, we have opted not to simulate $\nu$DM interactions within the galactic halo; therefore, our treatment is on the conservative side. However, this kind of interactions in the galactic halo would generate interesting signatures, such as a non-isotropic flux at Earth due to the Solar System being off-center of the DM halo.

For a detailed description of neutrino creation and annihilation processes within this model, we present in Appendix~\ref{app:xsec_nuDM} the differential cross section for these scatterings. Since for the parameter values relevant for our setup, the cross section is resonant only in a very narrow energy window unresolvable by GRAND, we will separately consider two regimes of interest: the parameter space region in which the center-of-mass energy is well above the resonance condition $s=2E_{\nu}m_S\gg m_\chi^2$ (i.e. light mediator) and the region in which $s=2E_{\nu} m_S\ll m_\chi^2$ (i.e. heavy mediator). In these two regimes, the interaction rate $\Gamma_{\nu S}$ reads
\begin{equation}
    \Gamma_{\nu S}=n_S\sigma_{\nu S}+n_{\bar S}\sigma_{\nu \bar{S}}=\tfrac12\rho_S
 \begin{cases} 
     \displaystyle  \frac{y^4}{4\pi \, m_\chi^4} E_\nu& E_\nu\ll \frac{m_\chi^2}{2m_S} \, ,\\
     \displaystyle \frac{y^4 \left(2\log\left(\frac{2E_\nu m_S}{m_\chi^2}\right)-1\right)}{64\pi \,  m_S^2 E_\nu} &  E_\nu\gg \frac{m_\chi^2}{2m_S}\, ,
   \end{cases}
   \label{eq:nuDM_int_rate}
\end{equation}
where, for simplicity, we have dropped the mass indices in the coupling and assumed equal number densities for $S$ and $\bar S$. Notice how, in the heavy mediator regime, the interaction rate is independent of the scalar DM mass $m_S$, while in the light mediator regime, $\Gamma_{\nu S}$ has a very mild logarithmic dependence on the fermion mediator mass $m_\chi$.

\begin{figure}[h!]
    \centering
    \includegraphics[width=0.9\textwidth]{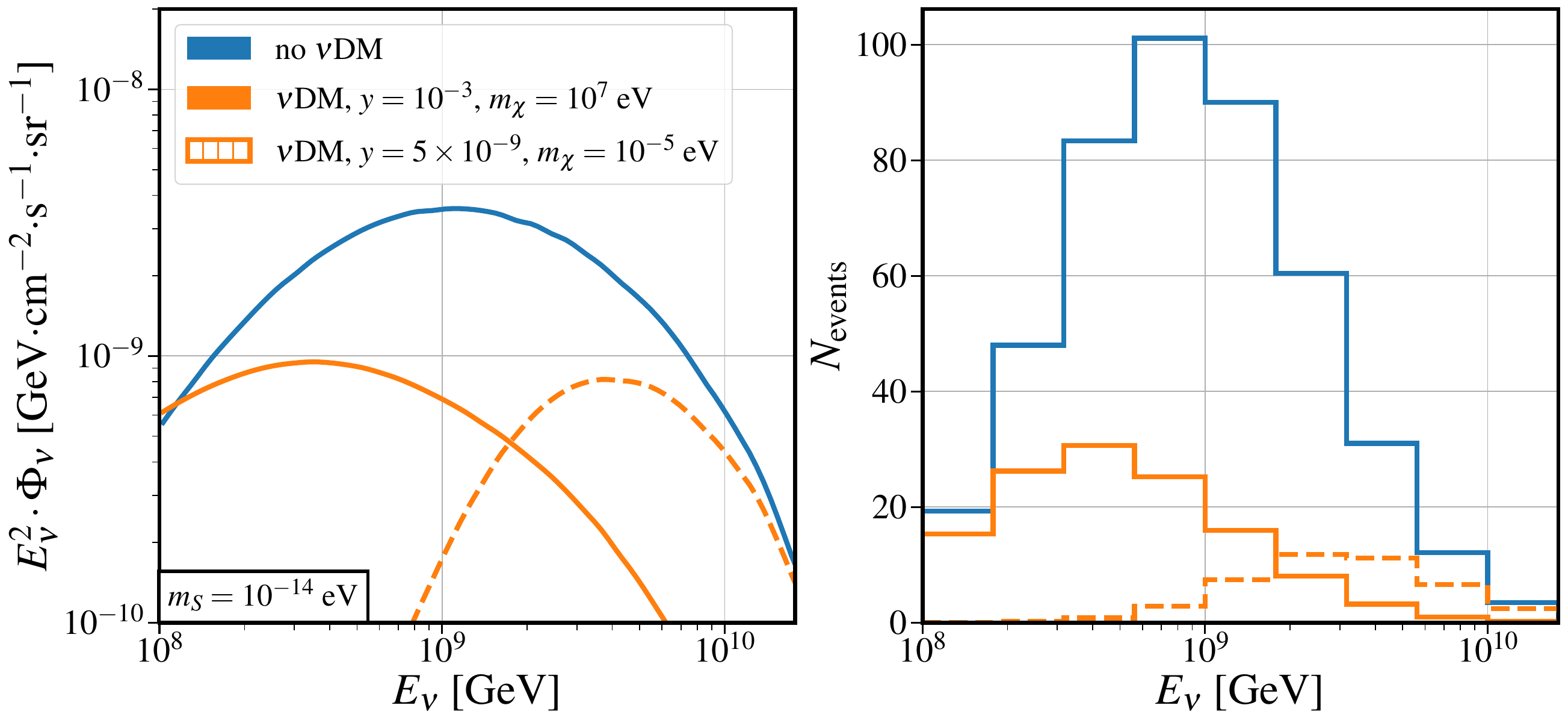}
    \caption{On the left panel we present the single-flavor cosmogenic neutrino flux at the Earth for $\gamma =3, m=3$ and $E_{\rm max}= 250$ EeV without $\nu$DM interaction (blue) and with $\nu$DM interaction for $y = 10^{-3}$ and $m_\chi= 10^7$ eV (solid orange) and for  $y = 5 \times 10^{-9}$ and $m_\chi= 10^{-5}$ eV (dashed orange), both for $m_S=10^{-14}$ eV. On the right panel we have the corresponding event distributions at GRAND (10-year exposure).}
    \label{fig:example_nuDM}
\end{figure}

Unlike the $\nu$SI scenario, in which the spectral dip originating from the resonance was crucial for the sensitivity, this scenario does not rely on this feature due to two reasons: first of all, the resonance is too narrow to be resolved and secondly, due to the DM being ultralight, the very large number densities greatly enhance the interaction rate such that, even for small couplings and non-resonant energies, the flux can still be greatly modified. In Fig.~\ref{fig:example_nuDM} we show the effect of $\nu$DM interactions in the cosmogenic neutrino flux for the two regimes of Eq.~\eqref{eq:nuDM_int_rate}. Due to the different energy dependence of the interaction rate, the suppression is more pronounced towards lower or higher energies, depending on the regime. It is evident that the effect of $\nu$DM interactions does not generally lead to pronounced spectral features, such that the shape of the modified flux might be somewhat degenerate with astrophysical parameters. However, even for relatively small couplings, the flux suppression induced by $\nu$DM interactions cannot be offset by even the most optimistic astrophysical assumptions (e.g. a very strong source evolution). 
Therefore, the sensitivity to this scenario does not rely on resolving specific spectral features and simply hinges on a positive detection of the cosmogenic flux, which would allow us to rule out parts of the parameter space in which $\nu$DM interactions opaque the Universe to cosmogenic neutrinos. This in turn means that GRAND's exclusion power will not be sizably reduced even if the flux measurement is statistically limited, as only a handful of events can already place robust constraints.

We compute the cosmogenic neutrino flux at Earth in the presence of $\nu$DM by solving the transport equation using a modified version of the \texttt{nuSIprop}~\cite{Esteban:2021tub} code employed in the $\nu$SI analysis, where we have replaced the collision terms of the $\nu$SI setup by our own computations in the $\nu$DM scenario, as well as replaced the C$\nu$B number density by the cosmological DM number density. For some choice of astrophysical parameters, the cosmogenic flux at Earth can then be calculated by specifying the coupling $y_\tau$, the DM mass $m_S$, and the mediator mass $m_\chi$.

\section{Analysis and Results}
\label{sec:ana}

\begin{figure}[h!]
    \centering
    \includegraphics[width=0.9\textwidth]{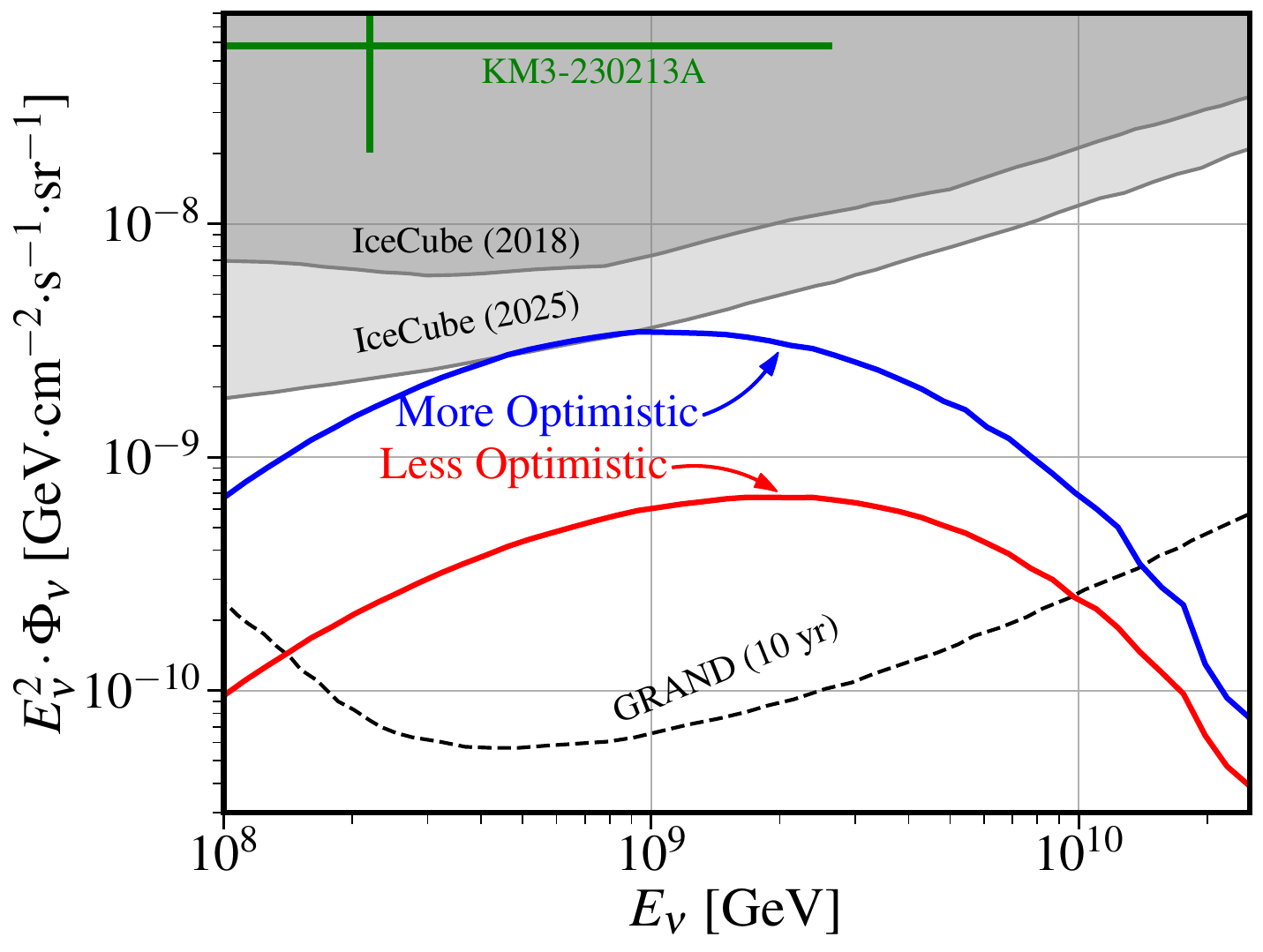}
    \caption{Single-flavor cosmogenic flux at GRAND for the two benchmark values of the astrophysical parameters, $\gamma_{A,B}=2.5$, $E_{{\rm max},A,B}=250$ EeV and $m_A$=3 (more optimistic choice, in blue) and $m_B=0$ (less optimistic choice, in red). We also show the two IceCube upper limits at 90\% C.L. (dark and light gray-shaded regions)~\cite{IceCube:2018fhm,IceCube:2025ezc} and GRAND's  projected 10-year sensitivity curve (dashed line)~\cite{GRAND:2018iaj}. We also show as a reference the 
    data point KM3-230213A~\cite{KM3NeT:2025npi} with $1\sigma$ uncertainties in its normalisation.}
    \label{fig:BenchmarkFluxes}
\end{figure}

In order to evaluate GRAND's sensitivity to
the BSM models we discussed in section~\ref{sec:BSM},  we generate mock data under the hypothesis of SM physics.  We will do this by computing the
cosmogenic neutrino spectrum for two benchmark values of the astrophysical parameters ($\gamma$, $E_{\rm max}$ and $m$): a more optimistic choice, which corresponds to  $\gamma_A=2.5$, $E_{{\rm max},A}=250$ EeV and $m_A=3$ and a less optimistic choice, which corresponds to  $\gamma_B=2.5$, $E_{{\rm max},B}=250$ EeV and $m_B=0$.  
Notice that, since the source composition and evolution are 
degenerate~(see last panel of Fig.~\ref{fig:varying-astrophysics}), these benchmark scenarios 
can be viewed as representative of cases in which the sources have heavier compositions and a stronger source evolution. We note, however, that many astrophysical models~\cite{Murase:2007yt,Aloisio:2015ega,Fang:2017zjf,AlvesBatista:2018zui,Alves:2025xul} predict a flux well below the sensitivity of GRAND. 
Nevertheless, the neutrino flux at EeV energies is highly uncertain; it may vary about 2 orders of magnitude depending on the evolution of the EECR sources and mass composition~\cite{Moller:2018isk,AlvesBatista:2018zui}.
There may be additional components of $E_\nu \gtrsim 1$ EeV 
neutrinos produced by largely unconstrained sources~\cite{Righi:2020ufi,Rodrigues:2020pli,Ackermann:2022rqc}.
In Fig.~\ref{fig:BenchmarkFluxes} we show the flux predictions for our two benchmark points as a function of the neutrino energy in the range $(10^8-3\times 10^{10})$~GeV. Notice how our more optimistic benchmark scenario is already in slight tension with the constraints from the recent IceCube cosmogenic search~\cite{IceCube:2025ezc}. However, the  data point KM3-230213A may be an indication that the true cosmogenic flux may be more  consistent with our more optimistic scenario, although this interpretation is in tension with IceCube's non-observation of similar events~\cite{Li:2025tqf}.

We then compute the prediction for the number of $(\nu_\tau+\bar \nu_\tau)$ events in each of the 9 energy bins we consider: $\hat{N_k}=\hat{N_k}(\gamma_{A,B},E_{{\rm max},A,B},m_{A,B})$ is the standard mock data prediction for one of the benchmark values for the cosmogenic spectrum and $N_k(\theta,\gamma,E_{\rm max},m)$ is the  corresponding prediction for a given BSM scenario which will also depend on $\theta$, the parameters of the model, as well as the flux parameters $\gamma$, $E_{\rm max}$ and $m$. 

The sensitivity power of GRAND for each BSM  model is then quantified by means of a test statistic, where we compare the number of events
$N_k\equiv N_k\left(\mathbf{\theta},\gamma,m\right)$ predicted for a BSM model at a given 
value of the model parameters $\mathbf{\theta}$, with the mock data value  $\hat{N}_k$. Our test statistic is 
defined by the Poisson log-likelihood 
\begin{equation}
\begin{split}
    \lambda(\mathbf{\theta},\gamma,E_{\rm max},m)&=-2\log \mathcal{L}(\mathbf{\theta},\gamma, E_{\rm max},m)\\&=2\sum_{k=1}^9\left[N_k(\mathbf{\theta},\gamma,E_{\rm max},m)-\hat{N}_k+\hat{N}_k\log\frac{\hat{N}_k}{N_k(\mathbf{\theta},\gamma,E_{\rm max},m)}\right]\, ,
\end{split}
    \label{eq:test_statistic}
\end{equation}
so GRAND's sensitivity  for fixed values of the BSM parameters $\mathbf{\theta}$ will be given by the value of $\lambda$ after minimizing over the three astrophysical parameters. We will vary them in the range\footnote{When testing mock data generated with $m=3$, we restrict ourselves to varying $m\in [2,4]$. Instead, when the mock data was generated with $m=0$, the chosen interval for the fit is $m\in[-1,1]$. These intervals are wide enough to take into account the degeneracies with the BSM parameters while being sufficiently narrow for an efficient scanning.} $m \in [-1.0,4.0]$, $\gamma \in [2.0,3.0]$ and $E_{\rm max}\in[10^2,10^5]$ EeV. This procedure allows us to construct the experiment's sensitivity region at a certain confidence level for the  parameters of each BSM model.

An important observation is the fact that, while changing the source evolution parameter $m$ amounts to a reweighting of the simulated events, changing $\gamma$ and $E_{\rm max}$ requires launching a new simulation, as the cosmogenic flux depends on these quantities in a non-parametric fashion. Therefore, in order to vary these two astrophysical parameters within their aforementioned ranges, we first generate several spectrum templates for different choices of $\gamma$ and $E_{\rm max}$ and then we proceed to scan the parameter space $\theta$ of the BSM setup along with the source evolution $m$. For each point in the scan, we compute the test statistic as in Eq.~\eqref{eq:test_statistic} for all of the spectrum templates, keeping the smallest value, i.e., we profile the test statistic over $\gamma$ and $E_{\rm max}$. After a subsequent profiling over $m$, we compute the sensitivity of the profiled test statistic $\lambda(\theta)$ under the assumption that it is distributed as a $\chi^2$ with degrees of freedom equal to the dimension of $\theta$.

\subsection*{$\nu$SI results}
\begin{figure}[h!]
    \centering
    \includegraphics[width=0.9\textwidth]{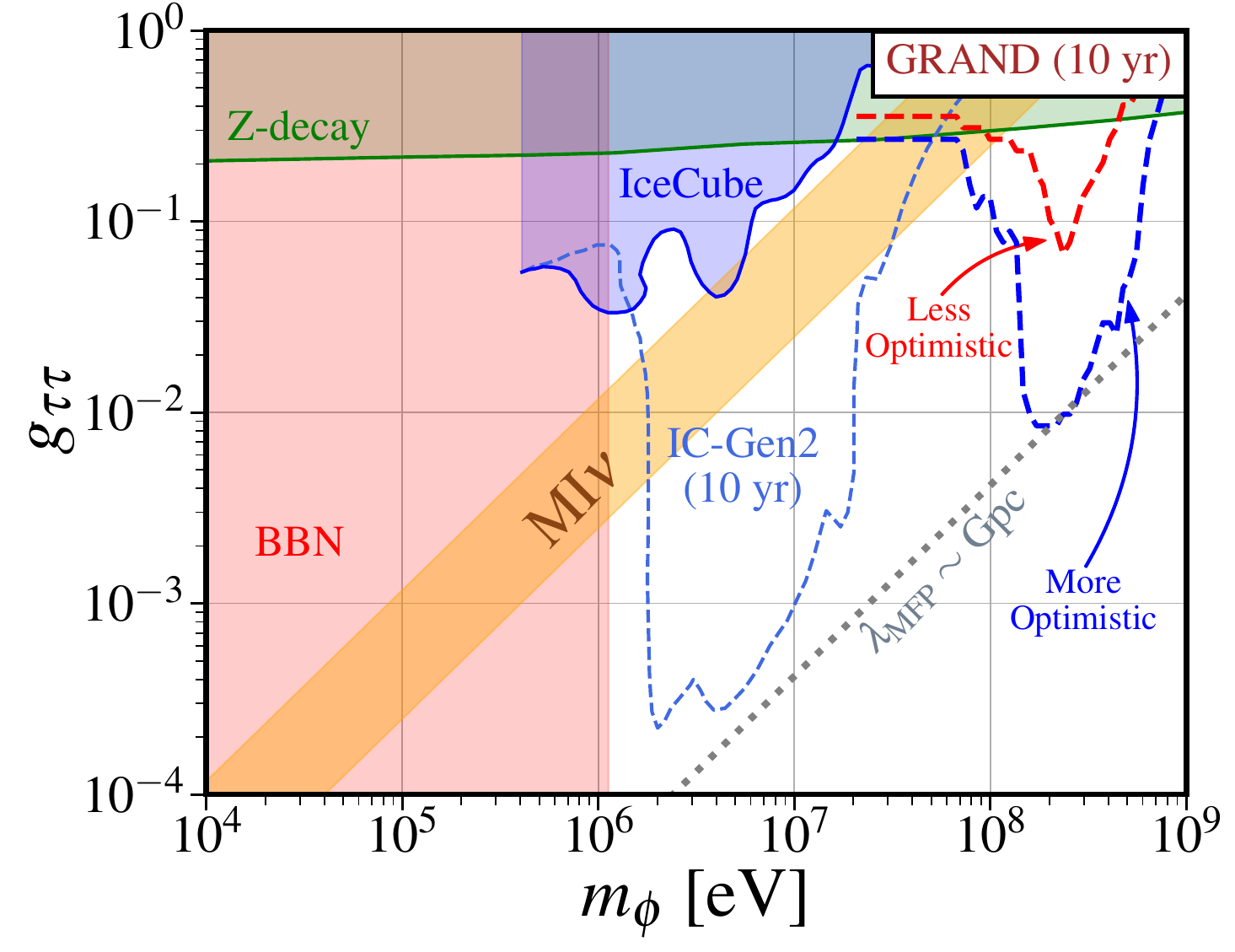}
    \caption{Future sensitivity of GRAND (10-year exposure)  on the $\nu$SI parameter space $g_{\tau \tau}$ versus $m_\phi$. The dashed blue (red) curve shows the sensitivity of GRAND for the more (less) optimistic benchmark values of the astrophysical parameters.
    We also show the $Z$ decay (green) and BBN (red) 95\% CL bounds as well as the constraints from IceCube HESE 7.5-year data (blue) and the IceCube-Gen2 10-year sensitivity curve, which are taken from Ref.~\cite{Esteban:2021tub}. For reference, we present in yellow the region where the tension in the Hubble parameter can be alleviated by $\nu$SI~\cite{Kreisch_2020}, labeled MI$\nu$. 
    }
    \label{fig:nuSI_bounds}
\end{figure}

Let's start the discussion of our results with the $\nu$SI model. 
There are already several limits on these lepton number violating scalar neutrino couplings. Let's start from laboratory bounds coming from meson $M$ decays,
i.e. from 
$M \to \ell_\alpha \nu_\beta \phi$
with $m_\phi < m_M-m_{\ell_\alpha}$.
In the lower mass range, kaons decays 
are the main reason for the limits
$(\sum_{\alpha} \vert g_\phi^{e \alpha}\vert^2)^{1/2} \lesssim 7 \times 10^{-3}$, for $m_\phi \lesssim 
200$  MeV, and  $(\sum_{\alpha} \vert g_\phi^{\mu \alpha}\vert^2)^{1/2} \lesssim 2.2 \times 10^{-2}$, for $m_\phi \lesssim 100$ MeV~\cite{Pasquini:2015fjv}.
In the range $0.1< m_\phi/{\rm GeV}< 2$
limits on $\vert g_\phi^{\alpha \beta}\vert$ from $M=K$, $D_s$ and $B$ 
leptonic decays as well as from  
invisible $Z\to\nu_\alpha\nu_\beta\phi$ decays are much weaker~\cite{deGouvea:2019qaz}. 
There are also limits from $\tau \to \ell_{\alpha} \nu_\alpha \nu_\tau \phi$ on $(\sum_{\alpha} \vert g_\phi^{\tau \alpha}\vert^2)^{1/2} < {\cal O}(1)$, for $m_\phi \lesssim 0.75$ MeV~\cite{Lessa:2007up,deGouvea:2019qaz}.
Neutrino self-interaction
can also cause neutrinoless double beta decay as $(A,Z) \to (A,Z+2) + 2e^- + \phi$, provided that $\phi$ is lighter than the $Q$-value of the decaying nucleus. Thus KamLAND-Zen~\cite{KamLAND-Zen:2012uen} and EXO~\cite{Kharusi:2021jez} can set the  limits 
$\vert g_\phi^{ee}\vert \lesssim 10^{-3} \, (10^{-5})$
 for $m_\phi \lesssim 2 \, (0.2)$ MeV~\cite{Blum:2018ljv}. 
 Coherent neutrino scattering can impose 
 bounds on couplings of the order $10^{-4}$, if $\phi$ couples to quarks and couplings to neutrinos are taken to be flavor universal~\cite{Farzan:2018gtr,AtzoriCorona:2022moj}.
 Luminosity and deleptonization constraints imposed on the observation of SN1987A exclude the 
 range $2.1 \times 10^{-9} \lesssim \vert g_\phi^{ee}\vert \times (m_\phi/{\rm MeV}) \lesssim 1.6 \times 10^{-6}$, 
 $5.5 \times 10^{-9} \lesssim \vert g_\phi^{\mu\mu,\tau \tau}\vert \times (m_\phi/{\rm MeV}) \lesssim 1.1 \times 10^{-6}$ and $2.3 \times 10^{-8} \lesssim \vert g_\phi^{e\mu,e \tau}\vert \times (m_\phi/{\rm MeV}) \lesssim 6.6 \times 10^{-7}$, for $m_\phi \in [10,500]$ MeV~\cite{Heurtier:2016otg}. However, smaller couplings can be constrained from the non-observation of $\mathcal{O}(100 \,{\rm MeV})$ neutrinos at Kamiokande-II and IMB during SN1987A~\cite{Fiorillo:2022cdq}. In particular, assuming a flavor universal coupling, Ref.~\cite{Fiorillo:2022cdq} found $10^{-7}\lesssim g_\phi\times(m_\phi/{\rm MeV})\lesssim 10^{-9}$ for $100\,{\rm eV}\lesssim m_\phi\lesssim 100\,{\rm MeV}$. If sterile neutrinos are also considered, this interaction can also be constrained via SN cooling arguments (see Ref.~\cite{Chen:2022kal}).
 Bounds from Big Bang Nucleosynthesis also exist, but they only apply for $m_\phi \lesssim 1$ MeV~\cite{Huang:2017egl,Blinov:2019gcj}. It is also possible to probe $\nu$SI by the observation of high energy neutrino fluxes~\cite{Esteban:2021tub,Ioka:2014kca,Murase:2019xqi}. Lastly, it has been shown that sufficiently strong $\nu$SI may affect the CMB power spectrum and partially alleviate the Hubble tension~\cite{Kreisch_2020}. In Fig.~\ref{fig:nuSI_bounds}, along with the different constraints, we show the region of parameter space, dubbed MI$\nu$ (Moderately Interacting Neutrinos), where this solution can be achieved. However, it should be noted that, while CMB and BAO data do showcase a preference for $\nu$SI, it has been recently shown that this preference goes way once the information from the full-shape analysis of the matter power spectrum is added~\cite{Poudou:2025qcx}.

Our results for $\nu$SI are shown in Fig.~\ref{fig:nuSI_bounds}.
In this case $\theta\equiv \{ g_{\tau\tau}, m_\phi \}$, 
which we vary in the range $g_{\tau\tau} \in [10^{-3},1]$
and $m_\phi \in [10^7,10^9]$ eV. In this figure,  we  show the sensitivity we derived for GRAND (colored dashed lines) with the current constraints (colored solid lines) from other experiments \cite{Esteban:2021tub}. We have also included the prospective sensitivity of the Cherenkov part of IceCube-Gen2~\cite{Esteban:2021tub} as a reference. We can see that IceCube-Gen2 has better sensitivity for the coupling $g_{\tau\tau}$
than GRAND's; however, for a lighter mediator, i.e., $m_\phi \sim $ a few MeV to a few tens of MeV. This is because IceCube-Gen2, using the Cherenkov effect, is expected to detect lower energy neutrinos. Moreover, as the resonant cross section $\sigma_{\text{res}}\sim1/m_{\phi}^2$ is larger for lighter mediators, the imprints of $\nu$SI in the IceCube-Gen2 spectrum are easier to detect, i.e., they lead to more pronounced spectral features. This aspect, combined with the superior energy resolution of Gen2 compared to GRAND, leads to the former being able to set stronger bounds 
on $g_{\tau\tau}$ by more than one order of magnitude, even when Gen2's statistics can potentially be comparable to GRAND's. We also plot as a gray dotted line the parameter space in which the mean free path, estimated as $\lambda^{-1}_{\rm MFP}\sim \frac{g_{\tau\tau}^2}{m_\phi^2}n_\nu$ is equal to the typical $\sim h_0^{-1}$ ($\sim $ Gpc) baselines traversed by cosmogenic neutrinos. This serves as a rough sensitivity limit for high-energy neutrino observatories, as couplings below the gray dotted line would barely produce any modification on high-energy extragalactic neutrinos. As pointed out in Ref.~\cite{Esteban:2021tub}, IceCube-Gen2 will have enough constraining power to probe this sensitivity limit. Remarkably, also GRAND (under optimistic flux assumptions) will also be able to probe along this maximum sensitivity line.

However, Fig.~\ref{fig:nuSI_bounds} also shows GRAND's complementarity to other experiments, as it can potentially constrain a previously unexplored region of the parameter space. The shape of GRAND's sensitivity can be easily understood. When the mediator is too light, the spectral dip induced by the resonant enhancement of the cross section falls at neutrino energies below GRAND's sensitivity threshold, which is the reason why the constraint sharply drops for $m_\phi\lesssim 100$ MeV. On the contrary, even when the mass of the mediator is too large such that the dip from the resonant cross section  falls above GRAND's energy sensitivity, cosmogenic neutrinos get down-scattered to lower energies, thus  producing a bump in the spectrum to which GRAND can still be somewhat sensitive. For this reason, the sensitivity curve falls in a slightly smoother way as one moves to larger mediator masses.

\subsection*{PD$\nu$ results}
\begin{figure}[h!]
    \centering
    \includegraphics[width=0.9\linewidth]{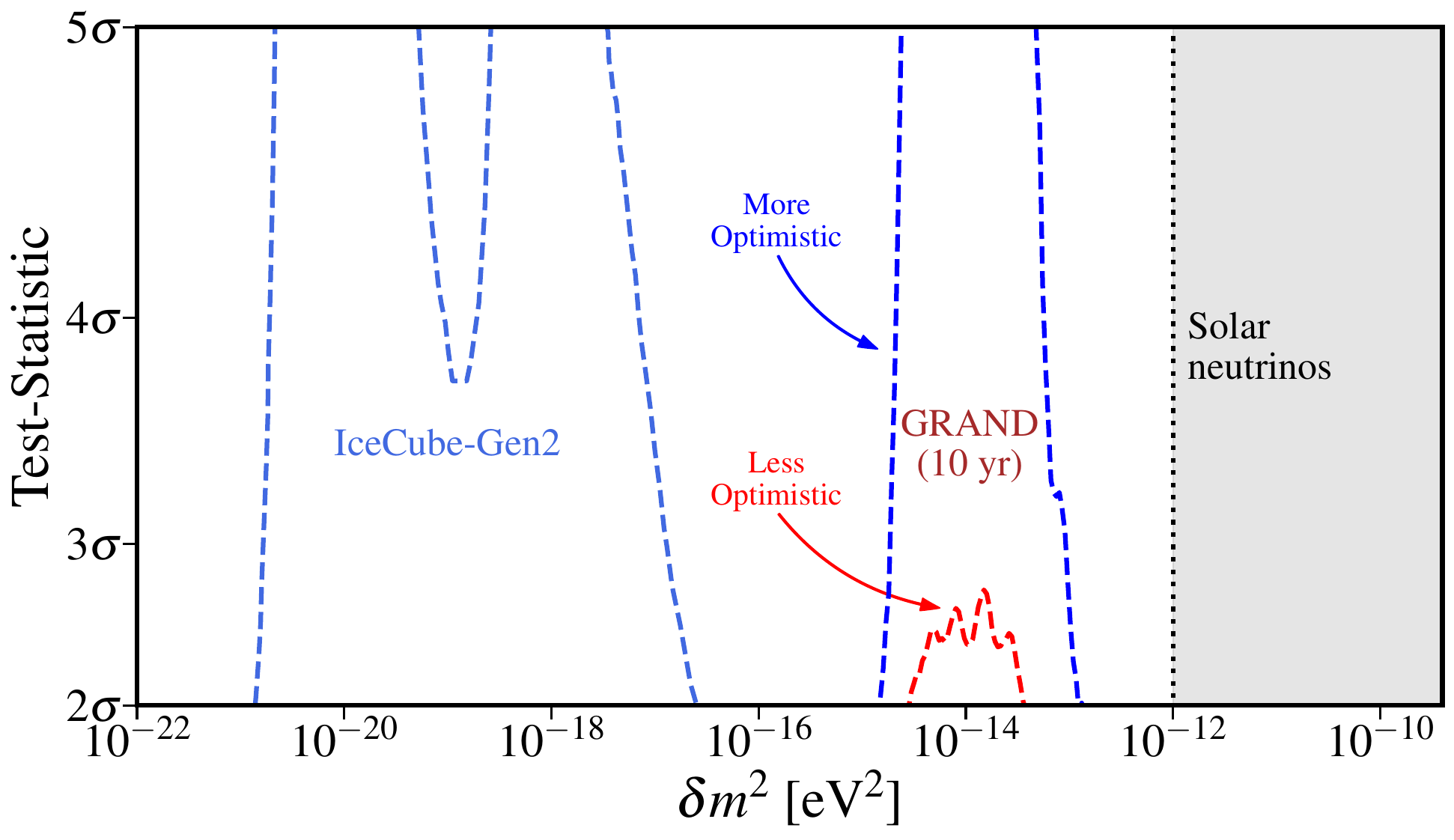}
    \caption{GRAND's sensitivity to  $\delta m^2$ for the more (less) optimistic cosmogenic flux is show in blue (red). We also show the region already excluded by solar neutrinos~\cite{Chen:2022zts} and the region that can be probed in the future by IceCube-Gen2~\cite{Carloni:2022cqz}.   }
    \label{fig:pseudodirac_exclusion}
\end{figure}

Next let us examine the PD$\nu$ model. In this case, there are also several limits on the mass splitting $\delta m^2$
from different experimental data. Solar neutrinos constrain
the mass splitting to be in the region $\delta m^2\lesssim 10^{-12}$ eV$^2$~\cite{Chen:2022zts,Ansarifard:2022kvy} while atmospheric 
neutrinos provide the much weaker bound $\delta m^2\lesssim 10^{-4}$ eV$^2$~\cite{Beacom:2003eu}. An analysis of neutrinos 
from SN1987A can exclude smaller splittings but for the narrow range $\delta m^2 \in [2.55,3.01] \times 10^{-20}$ eV$^2$~\cite{Martinez-Soler:2021unz}. IceCube data recently allowed for limits to be set in the range $\delta m^2 \in [10^{-21},10^{-16}]$ eV$^2$~\cite{Carloni:2022cqz,Rink:2022nvw,Dixit:2024ldv,Carloni:2025dhv}.
In Fig.~\ref{fig:pseudodirac_exclusion} we show GRAND's complementary sensitivity to the range $\delta m^2 \in [10^{-15},10^{-13}]$ eV$^2$, a range 
not probed by previous experiments. For comparison, we plot the forecasted exclusion power of IceCube-Gen2~\cite{Carloni:2022cqz}, as well as the excluded region from solar neutrinos. Indeed, thanks to its higher energies, which suppress the oscillation frequency, GRAND can probe values of $\delta m^2$ for which active-sterile oscillations would be averaged out in IceCube. The shape of the sensitivity region can be easily understood: the $\delta m^2$ interval to which GRAND is sensitive corresponds to the values for which the first oscillation dip falls within its energy range. Instead, for smaller mass splittings, oscillations do not have the time to develop, while for larger $\delta m^2$ oscillations average out and amount to a factor two suppression of the flux normalization, which cannot be disentangled from astrophysics.

Also note the strong degradation in the sensitivity from the more optimistic scenario to the less optimistic one. Indeed, in the latter case, the number of events per bin is $\sim 10$ or lower, which does not yield strong discriminating power against a signal that amounts to depletion in some small energy window, as statistical fluctuations can partially mimic such a signature.

\subsection*{$\nu$DM results}
We will now discuss the $\nu$DM model. This kind of interaction can be constrained from measurements of neutrino fluxes that traverse galactic or even cosmological scales. In particular, bounds on the same type of interaction considered in our work were derived in Ref.~\cite{Sen:2023uga} employing the IceCube measurement of neutrinos from the blazar TXS 0506+056 and from SN1987A. The mere observation of neutrinos coming from sources at known distances allows us to constrain the strength of $\nu$DM interactions by requiring the Universe to not be opaque for such neutrinos. It is also possible to constrain $\nu$DM from observations of the matter power spectrum at small scales from the Ly-$\alpha$ forest, the Milky Way's satellites, or even the CMB, since diffusion damping within the neutrino-dark matter coupled fluid in the early Universe suppresses power below the damping scale~\cite{Bertoni:2014mva,Wilkinson:2014ksa,Escudero:2018thh,Brax:2023tvn,Akita:2023yga}.

\begin{figure}[h!]
    \centering
    \vglue -0.cm
    \includegraphics[width=0.9\linewidth]{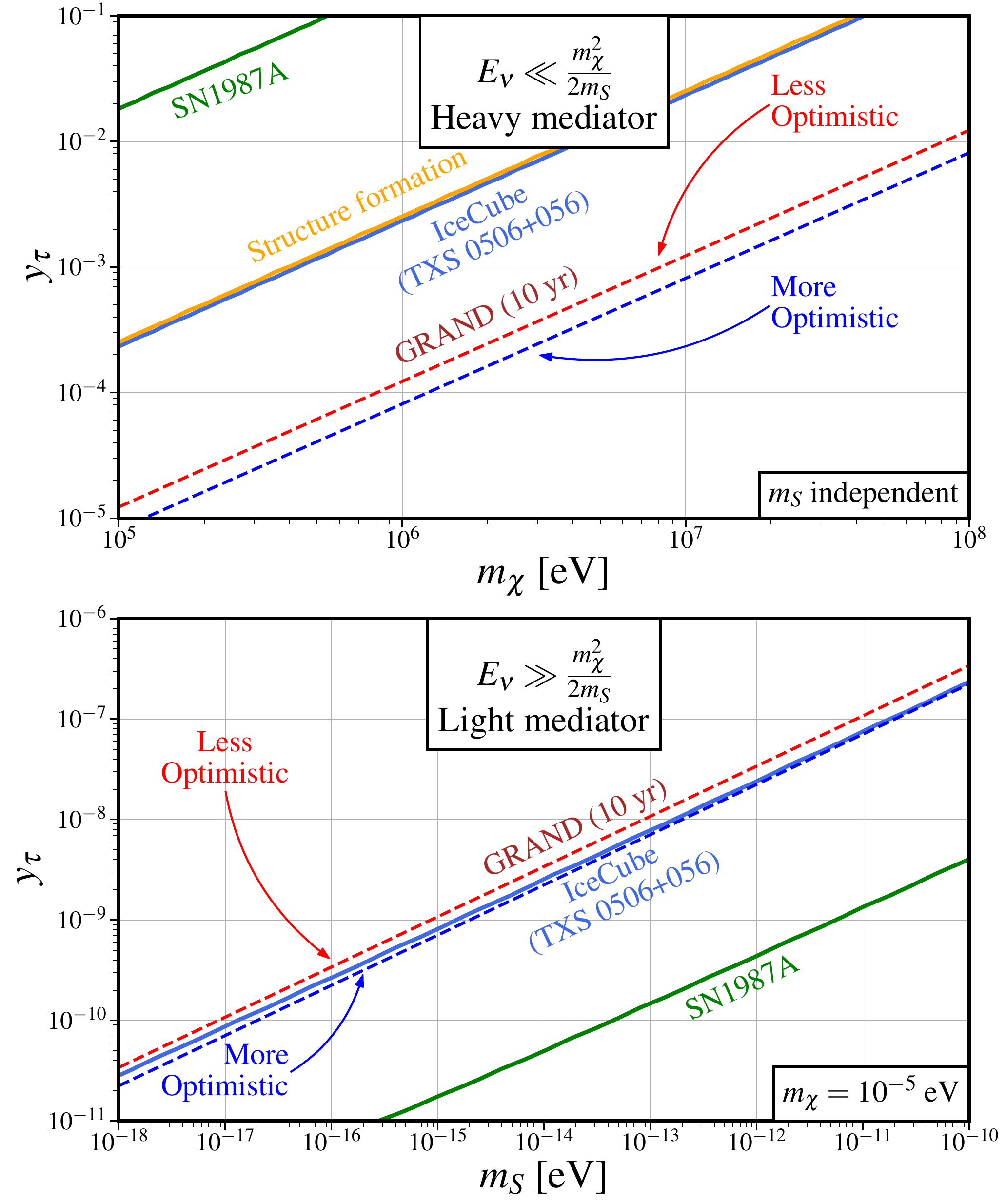}
    \caption{Here we show GRAND's 
    10 year $95\%$ CL sensitivity for the more (less) optimistic values of the astrophysical parameters in dashed blue (dashed red). In the upper (lower) panel we have the $E_\nu \ll m^2_\chi/2 m_S$ ($E_\nu \gg m^2_\chi/2 m_S$) regimes. We also show the limits from SN1987A (green), structure formatiom (yellow) and IceCube TXS 0506+056 (solid blue).}  
    \label{fig:nuDM_bounds}
\end{figure}

Our results for GRAND's sensitivity to $\nu$DM are summarized in Fig.~\ref{fig:nuDM_bounds}. Since we have separated this setup into two regimes (heavy and light mediators), we present our results in the two panels of Fig.~\ref{fig:nuDM_bounds}. Notice the different horizontal axis labels, since each of the two regimes probes very different parts of parameter space (see Eq.~\eqref{eq:nuDM_int_rate}). Indeed, as outlined in Section~\ref{sec:BSM}, when the neutrino energy is below resonance, the interaction rate is only given by the ratio $y/m_{\chi}$ and is independent of the DM mass $m_S$ (as long as the condition $E_\nu\gg\tfrac{m_\chi^2}{2m_S}$ is met), while in the regime above resonance, the interaction rate is mostly set by $y^2/m_S$, while retaining only a logarithmic dependence on the mediator mass $m_\chi$. We also plot the constraints from IceCube and SN1987A derived in Ref.~\cite{Sen:2023uga} for comparison. In the heavy mediator regime we also plot the bound arising from the observation of the matter power spectrum at small scales~\cite{Wilkinson:2014ksa,Escudero:2018thh} via the Lyman-$\alpha$ forest and the Milky Way's satellites. Due to the ultralightness of our DM candidate, the cross section grows linearly with temperature, in stark contrast with the $\propto T^0$ and $\propto T^2$ behaviors commonly adopted in the literature when analyzing $\nu$DM in cosmological contexts. Consequently, it is not entirely straightforward to recast the bounds available in the literature into our model. As a rough estimation, we evaluate our cross section at $\braket{E_\nu}=\tfrac{7\pi^4}{180 \zeta(3)}T$ with $T\sim 250$ keV, where typically bounds on $\propto T^2$ and temperature-independent cross sections equate~\cite{Wilkinson:2014ksa}. We then employ the bound on a $T$-independent cross section extracted in Ref.~\cite{Escudero:2018thh}. We do not plot the corresponding cosmological $\nu$DM bound for the regime above-resonance. The reason is twofold: the qualitatively different behavior with temperature ($\propto T^{-1}$) precludes a simple recasting of existing constraints. In addition, this kind of interaction will cross from the light to the heavy mediator regime at some point throughout cosmological evolution, which will render the bounds extracted from cosmology very sensitive to the location of this resonance crossing.

The shape of the constraints can be intuitively understood as corresponding to isocontours of the interaction rate (see Eq.~\ref{eq:nuDM_int_rate}). Moreover, notice how, in stark contrast with the previous BSM scenarios, there is little loss of sensitivity when going from the more optimistic to the less optimistic detection scenario. As previously discussed, this is a consequence of the fact that the sensitivity stems from the mere observation of a handful of neutrinos, allowing to exclude the region of parameter space in which the $\nu$DM interaction would exponentially suppress such neutrino flux. Regarding GRAND's exclusion power relative to the already existing constraints, the differences can be easily understood in terms of the different energy scales of the involved neutrinos. In the heavy mediator regime (Fig.~\ref{fig:nuDM_bounds} upper panel), the interaction rate~\eqref{eq:nuDM_int_rate} scales with $E_\nu$ and thus GRAND, with its higher energies, will be substantially more sensitive than IceCube and SN1987A, with $O(\rm 1\, PeV)$ and $O(\rm 10\,MeV)$ neutrinos, respectively. Although a dedicated cosmological analysis of this kind of $\nu$DM scenario is required, our naive recasting also indicates that cosmogenic neutrinos can place stronger bounds than structure formation. Conversely, in the light mediator regime, the situation is reversed, as the interaction rate scales as $1/E_\nu$, thus making SN1987A the best probe in this region of parameter space, as it involves the lowest neutrino energies. Notice how TXS 0506+056 sensitivity is comparable with GRAND's. Even though the energy scale of the neutrino from the blazar lies in the $\sim100$ TeV range, substantially below the cosmogenic neutrino scale, GRAND's sensitivity to high redshifts overcomes the suppression of the interaction rate due to higher DM densities.

\section{Final Conclusions}
\label{sec:conc}
The scattering of EECRs with photons of the CMB and of the EBL is expected to produce 
as a by-product neutrinos with energies of about 1 EeV, called cosmogenic neutrinos. 
The cosmogenic neutrino diffuse flux, however, is highly unknown due to a number of astrophysical uncertainties concerning the properties of the primary EECR sources.
Recently, the  observation of a near-horizontal neutrino event of about 0.1 EeV by KM3NeT~\cite{KM3NeT:2025npi}, which may 
constitute the first detection of cosmogenic neutrinos, seems to indicate that the cosmogenic neutrino flux may  well be within the reach of the future radio array of antennas, such as the one proposed by the GRAND project and IceCube-Gen2. This possibility represents a great opportunity to use cosmogenic neutrinos not only to learn about these extreme astrophysical sources and probe tau neutrino properties but also to investigate BSM scenarios.

By assuming that the cosmogenic neutrino flux 
is in fact within the reach of these future radio detectors, we investigate in our paper the sensitivity of GRAND to a few well-motivated BSM models.
GRAND proposes to use  radio pulse detectors to measure the cosmogenic $\nu_\tau/\bar \nu_\tau$ flux by means of   Earth-skimming EAS produced by  neutrinos that  interact close to the surface producing a $\tau$ lepton as they emerge from the Earth. 
Since the cosmogenic neutrino flux is plagued 
by astrophysical uncertainties,  
 we concentrate on models that can modify the 
 single-flavor ($\nu_\tau$+$\bar \nu_\tau$) diffuse cosmogenic spectrum at GRAND in a distinctive observable way, not easily mimicked by our assumptions about the astrophysical sources.

We model the cosmogenic neutrino flux 
by assuming a distribution of identical pure-proton sources with the same luminosity which extend up to redshift $z_{\rm max}=7$ and can be described by three parameters: 
the power-law injection spectrum spectral index $\gamma$, the maximal proton energy $E_{\rm max}$ and 
the source evolution parameter $m$. Because of known parameter degeneracies, this simple   parametrization can also simulate the cosmogenic spectrum 
for more complex chemical compositions.
We investigate two benchmark points ($A$ and $B$) for the 
cosmogenic flux: $\gamma_A=2.5$, $E_{{\rm max},A}= 250$ EeV and $m_A=3$ (more optimistic) 
and $\gamma_B=2.5$, $E_{{\rm max},B}=250$ EeV and $m_B=0$ (less optimistic).

We examine three BSM cases: $\nu$SI - neutrino self-interactions (cosmogenic neutrinos interacting with the C$\nu$B via a scalar portal), PD$\nu$ - pseudo-Dirac neutrinos (cosmogenic neutrinos are sensitive to neutrino active-sterile oscillations involving tiny mass squared  splittings)   and $\nu$DM (cosmogenic neutrinos scattering 
on ultra-light DM). 

We find that GRAND can 
be sensitive to a currently unconstrained parameter region of the 
$\nu$SI model. It can constrain the coupling in the range $10^{-2}\lesssim g_{\tau \tau}\lesssim 10^{-1}$  for $7 \times 10^7 \lesssim m_\phi/{\rm eV} \lesssim 7 \times 10^8$ in case the cosmogenic flux is consistent with benchmark $A$. This is complementary to what is expected to be covered by IceCube-Gen2.  

According to our study, the GRAND project  seems to be very sensitive to 
 PD$\nu$ for a mass squared splitting in the range  $10^{-15}\lesssim \delta m^2/{\rm eV^2} \lesssim 10^{-13}$, assuming a more optimistic cosmogenic flux.
This is a currently unconstrained region for $\delta m^2$ and again a region that is  complementary to the one expected to be accessible by IceCube-Gen2.

Finally, for the $\nu$DM model, contrary to the previous scenarios, the sensitivity of GRAND is similar for both 
the more optimistic and the less optimistic cases. In the case of the heavy mediator scenario, due to the higher energies involved, GRAND  can do substantially better than the current limits from SN1987A, IceCube and our recasting of bounds from structure formation.  For a light mediator, however, GRAND limits are at most similar to IceCube and much weaker than those obtained using SN1987A data.

To conclude, we have shown that as long as 
the one-flavor cosmogenic neutrino flux is 
consistent with our two benchmark points, 
it can be used to constrain BSM physics at GRAND in a region of the parameter space complementary to other data.

%%%%%%%%%%%%%%%%%%%%%%%%%%%%%%%%%%%%%%%%
%%%%%%%%%%%%%%%%%%%%%%%%%%%%%%%%%%%%%%%%
\section*{Acknowledgments}\label{sec:acknoledgment}
%%%%%%%%%%%%%%%%%%%%%%%%%%%%%%%%%%%%%%%%
%%%%%%%%%%%%%%%%%%%%%%%%%%%%%%%%%%%%%%%%

We thank Rafael Alves Batista, Miguel Escudero, Enrique Fernandez-Martinez and Gaetano di Marco for helpful discussions. This project has received support from the European Union’s Horizon 2020 research and innovation programme under the Marie Skłodowska-Curie grant agreement N$^\circ$~860881-HIDDeN and N$^\circ$~101086085-ASYMMETRY. L.P.S.~Leal is fully financially supported by FAPESP under Contracts No. 2021/02283-6 and No. 2023/12330-7. R.~Zukanovich Funchal is partially supported by FAPESP under Contract No. 2019/04837-9, and by  Conselho Nacional de Desenvolvimento Científico e Tecnológico (CNPq). DNT acknowledges support from the HPC-Hydra cluster at IFT. The work of DNT was supported by the Spanish MIU through the National Program FPU (grant number FPU20/05333). DNT would like to warmly thank the Department of Mathematical Physics at USP for their hospitality, where this project was started. L.P.S.~Leal would like to thank the IJCLab Theory Group for their hospitality.

\clearpage

\appendix

\section{On the propagation of Cosmic Rays}
\label{app:transport}
For an isotropic and homogeneous distribution of cosmic ray sources, emitting  particles of type $i$, 
we can define the number 
of particles per co-moving volume and 
per unit of energy $E$ at time $t$ as
\begin{align}
    n_i(t,E) \equiv \frac{dN_i(t)}{(1+z)^3 \, dA d\Omega dE}\, , 
\end{align}
the time evolution of $n_i(t,E)$
is governed by a set of 1-dimensional transport equations~\cite{Ahlers:2009rf,Ahlers:2012rz} 
\begin{eqnarray}
    \partial_t n_i(t,E)  &= &\partial_E(\tilde b_i (t,E) \, n_i(t,E) ) +{\cal L}_i(t,E) \nonumber \\ 
    &-&\Gamma_i \, n_i(t,E) +\sum_j \int dE' \, \frac{d \sigma_{ji}(E,E')}{dE} \, n_j(t,E')\, , 
    \label{eq:transp}
\end{eqnarray}
 where $E=E(t,\epsilon)$,
 $H(t) = h_0 \sqrt{\Omega_\Lambda + \Omega_m (1+z)^3}$  is the cosmic expansion rate following the $\Lambda$CDM model with the fractions of dark energy and matter density 
 today, $\Omega_\Lambda = 0.685$ and $\Omega_m = 0.315$, respectively. The 
 normalizing Hubble parameter is
 $h_0=67.3$ km s$^{-1}$ Mpc$^{-1}$~\cite{Planck:2018vyg}. 
 Let's discuss the meaning of the 
 terms on the  r.h.s. of Eq.~\eqref{eq:transp}.
 In the first term, we have $\tilde b_i(t,E)$, which describes
the energy loss due to redshift and other continuous processes.
The second
term, ${\cal L}_i$, corresponds to the luminosity density per co-moving volume of sources emitting particles $i$. 
The third term describes the annihilation of $i$-type particles through the interaction rate $\Gamma_i$, while the last term refers to the regeneration of particles $i$ by an interaction of the form $j \to i$. Note that $E(z,\epsilon)$ is 
the solution of
 \begin{align}
 \tilde b_i(z,E(z,\epsilon)) =-\frac{d E(z,\epsilon)}{dt} = - H(z) (1+z) \frac{d E(z,\epsilon)}{dz} = E(z,\epsilon) H(z) + b_i(z,E(z,\epsilon))\, ,
 \end{align}
 with the initial condition $E(0,\epsilon)=\epsilon$,
 i.e. it is the energy that a particle $i$ had at redshift $z$ if it is observed at the Earth as having energy $\epsilon$. To derive this identity, we have also used the relation $dt = -dz/((z+1)H(z))$. Note that the energy loss rate with time is written as a combination of $E(z,\epsilon) H(z)$, the energy loss due to redshift, and the rate of other continuous energy losses $b_i(z,E(z,\epsilon))$. 

In order to solve the transport equation, we define 
\begin{align}
\tilde n_i(t,E) \equiv (1+z) n_i(t,E)\, ,
\end{align}
to re-write Eq.~\eqref{eq:transp} as 
\begin{align}
    \frac{d \tilde n_{i}(z,E)}{dz} = - \frac{1}{H(z)}{\cal L}_i^{\rm eff}(z,E) 
    - \frac{[\partial_E b_i(z,E) - \Gamma_i(z,E)]} {(z+1)H(z)} \, \tilde n_{i}(z,E)\,, 
\end{align}
so that 
\begin{align}
 \tilde n_i(z,E) = \int_{z}^{z_{\rm max}}
dz  \,  \exp \left [ \int_0^z \frac{\partial_E b_i(z',E(z',\epsilon)) - \Gamma_i(z',E(z',\epsilon))} {(z'+1)H(z')}  dz' \right] \frac{1}{H(z)}
{\cal L}_i^{\rm eff}(z,E(z,\epsilon))\, ,
\end{align}
where  ${\cal L}_i^{\rm eff}$ is defined as 
\begin{align}
{\cal L}_i^{\rm eff}(z,E(z,\epsilon)) =  {\cal L}_i(z, E(z,\epsilon)) +  
\sum_j \int dE' \, \frac{d \sigma_{ji}(E,E')}{dE} \, \frac{\tilde n_j(z,E')}{(1+z)}\, .     
\end{align}

In our study ${\cal L}_i(z,E)={\rm SE}(z,m) \, {\cal L}_i(0,E)$ and ${\cal L}_i(0,E) \propto dN_p/dE_p$.
The flux of particle $i$ at Earth (per unit of area, solid angle and energy) is then 
\begin{align}
    \phi_i(\epsilon) = \frac{1}{4\pi} \tilde n_i(0, E(0, \epsilon)).
\end{align}

\section{Neutrino-Ultralight Dark Matter Cross Section}
\label{app:xsec_nuDM}

For scalar dark matter, the interactions are given by

\begin{equation}
    \mathcal{L}\supset - \,y_k \,\overline{\chi_R} \,\nu_{kL}\,S + \,y_k^* \,\overline{\nu}_{kL} \,\chi_R \,S^*,
\end{equation}
where $i=1,2,3$ represent mass indices. In order to compute the neutrino flux arriving at the Earth, we need to propagate the neutrino number density considering their interactions with dark matter. This is done by employing the transport equation (see Appendix~\ref{app:transport}), where the depletion of $n_i$ is governed by
\begin{align}
    \Gamma_i&= \sum_{S^\prime=S,\bar S} \sum_{j}  n_{S^\prime}\,\sigma_{i S^\prime \to j S^\prime},
\end{align}
while the production of $i$ neutrinos is described by
\begin{align}
    \dfrac{d\sigma_{ji}}{dE}&= \sum_{S^\prime=S,\bar S} n_{S ^\prime}\dfrac{d\sigma_{j S^\prime\to i S^\prime}}{dE},
\end{align}
where $E$ denotes the energy of the $i$-type neutrino.

The annihilation cross section is given by
\begin{equation}
    \sigma_{i S(\bar S)\to j S(\bar S)} = \dfrac{\vert y_i\vert^2 \vert y_j\vert^2}{64\pi\,s}  \left\{s\left(\dfrac{-s^2+5\,m_\chi^2\, s-2m_\chi^4}{ \,( s - m_\chi^2)^2(s + m_\chi^2)}\right) + 2\log\left(\dfrac{s + m_\chi^2}{m_\chi^2}\right)\right\} ,\nonumber
\end{equation}
while the differential cross section for $i$-type neutrino creation can be written as
\begin{align}
    \dfrac{d\sigma_{j S(\bar S)\to i S(\bar S)}}{dE_i}&= -2 m_S\dfrac{d\sigma_{j S(\bar S)\to i S(\bar S)}}{dt} = \dfrac{m_S \,t \,\vert y_i\vert^2 \vert y_j\vert^2 (2 m_\chi^4 - 2m_\chi^2(s+t)+s^2+t^2)}{16\pi \,s \,( s - m_\chi^2)^2( t - m_\chi^2)^2},\nonumber
\end{align}
where $s$ is the squared center-of-mass energy and $t = (p_{S}^{\rm ini.} - p_i)^2$.

------------------------------------------------------------------------------------------------------------

%%%%%%%%%%%%%%%%%%%%%%%%%%%%%%%%%%%%%%%%

\bibliographystyle{JHEP} 
\bibliography{biblio}% Produces the bibliography via BibTeX.

\providecommand{\href}[2]{#2}\begingroup\raggedright\begin{thebibliography}{100}

\bibitem{Super-Kamiokande:1998kpq}
{\bf Super-Kamiokande} Collaboration, Y.~Fukuda et~al., {\it {Evidence for
  oscillation of atmospheric neutrinos}},  {\em Phys. Rev. Lett.} {\bf 81}
  (1998) 1562--1567, [\href{http://arxiv.org/abs/hep-ex/9807003}{{\tt
  hep-ex/9807003}}].

\bibitem{IceCube:2013low}
{\bf IceCube} Collaboration, M.~G. Aartsen et~al., {\it {Evidence for
  High-Energy Extraterrestrial Neutrinos at the IceCube Detector}},  {\em
  Science} {\bf 342} (2013) 1242856,
  [\href{http://arxiv.org/abs/1311.5238}{{\tt arXiv:1311.5238}}].

\bibitem{Berezinsky:1970xj}
V.~S. Berezinsky and G.~T. Zatsepin, {\it {Cosmic neutrinos of superhigh
  energy}},  {\em Yad. Fiz.} {\bf 11} (1970) 200--205.

\bibitem{Stecker:1973sy}
F.~W. Stecker, {\it {Ultrahigh energy photons, electrons and neutrinos, the
  microwave background, and the universal cosmic ray hypothesis}},  {\em
  Astrophys. Space Sci.} {\bf 20} (1973) 47--57.

\bibitem{Hill:1983xs}
C.~T. Hill and D.~N. Schramm, {\it {Ultrahigh-Energy Cosmic Ray Neutrinos}},
  {\em Phys. Lett. B} {\bf 131} (1983) 247.

\bibitem{AlvesBatista:2018zui}
R.~Alves~Batista, R.~M. de~Almeida, B.~Lago, and K.~Kotera, {\it {Cosmogenic
  photon and neutrino fluxes in the Auger era}},  {\em JCAP} {\bf 01} (2019)
  002, [\href{http://arxiv.org/abs/1806.10879}{{\tt arXiv:1806.10879}}].

\bibitem{KM3NeT:2025npi}
{\bf KM3NeT} Collaboration, S.~Aiello et~al., {\it {Observation of an
  ultra-high-energy cosmic neutrino with KM3NeT}},  {\em Nature} {\bf 638}
  (2025), no.~8050 376--382.

\bibitem{Ahlers:2012rz}
M.~Ahlers and F.~Halzen, {\it {Minimal Cosmogenic Neutrinos}},  {\em Phys. Rev.
  D} {\bf 86} (2012) 083010, [\href{http://arxiv.org/abs/1208.4181}{{\tt
  arXiv:1208.4181}}].

\bibitem{Aloisio:2015ega}
R.~Aloisio, D.~Boncioli, A.~di~Matteo, A.~F. Grillo, S.~Petrera, and
  F.~Salamida, {\it {Cosmogenic neutrinos and ultra-high energy cosmic ray
  models}},  {\em JCAP} {\bf 10} (2015) 006,
  [\href{http://arxiv.org/abs/1505.04020}{{\tt arXiv:1505.04020}}].

\bibitem{Ackermann:2022rqc}
M.~Ackermann et~al., {\it {High-energy and ultra-high-energy neutrinos: A
  Snowmass white paper}},  {\em JHEAp} {\bf 36} (2022) 55--110,
  [\href{http://arxiv.org/abs/2203.08096}{{\tt arXiv:2203.08096}}].

\bibitem{Denton:2020jft}
P.~B. Denton and Y.~Kini, {\it {Ultra-High-Energy Tau Neutrino Cross Sections
  with GRAND and POEMMA}},  {\em Phys. Rev. D} {\bf 102} (2020) 123019,
  [\href{http://arxiv.org/abs/2007.10334}{{\tt arXiv:2007.10334}}].

\bibitem{Valera:2022ylt}
V.~B. Valera, M.~Bustamante, and C.~Glaser, {\it {The ultra-high-energy
  neutrino-nucleon cross section: measurement forecasts for an era of cosmic
  EeV-neutrino discovery}},  {\em JHEP} {\bf 06} (2022) 105,
  [\href{http://arxiv.org/abs/2204.04237}{{\tt arXiv:2204.04237}}].

\bibitem{Huang:2021mki}
G.-y. Huang, S.~Jana, M.~Lindner, and W.~Rodejohann, {\it {Probing new physics
  at future tau neutrino telescopes}},  {\em JCAP} {\bf 02} (2022), no.~02 038,
  [\href{http://arxiv.org/abs/2112.09476}{{\tt arXiv:2112.09476}}].

\bibitem{Huang:2022pce}
G.-y. Huang, S.~Jana, M.~Lindner, and W.~Rodejohann, {\it {Probing heavy
  sterile neutrinos at neutrino telescopes via the dipole portal}},  {\em Phys.
  Lett. B} {\bf 840} (2023) 137842,
  [\href{http://arxiv.org/abs/2204.10347}{{\tt arXiv:2204.10347}}].

\bibitem{GarciaSoto:2022vlw}
A.~Garcia~Soto, D.~Garg, M.~H. Reno, and C.~A. Arg\"uelles, {\it {Probing
  quantum gravity with elastic interactions of ultrahigh-energy neutrinos}},
  {\em Phys. Rev. D} {\bf 107} (2023), no.~3 033009,
  [\href{http://arxiv.org/abs/2209.06282}{{\tt arXiv:2209.06282}}].

\bibitem{Heighton:2023qpg}
R.~Heighton, L.~Heurtier, and M.~Spannowsky, {\it {Hunting for neutral leptons
  with ultrahigh-energy neutrinos}},  {\em Phys. Rev. D} {\bf 108} (2023),
  no.~5 055009, [\href{http://arxiv.org/abs/2303.11352}{{\tt
  arXiv:2303.11352}}].

\bibitem{Kirk:2023fin}
M.~Kirk, S.~Okawa, and K.~Wu, {\it {A \ensuremath{\nu} window onto
  leptoquarks?}},  {\em JHEP} {\bf 12} (2023) 093,
  [\href{http://arxiv.org/abs/2307.11152}{{\tt arXiv:2307.11152}}].

\bibitem{ANITA:2021xxh}
{\bf ANITA} Collaboration, R.~Prechelt et~al., {\it {Analysis of a tau neutrino
  origin for the near-horizon air shower events observed by the fourth flight
  of the Antarctic Impulsive Transient Antenna}},  {\em Phys. Rev. D} {\bf 105}
  (2022), no.~4 042001, [\href{http://arxiv.org/abs/2112.07069}{{\tt
  arXiv:2112.07069}}].

\bibitem{ARA:2019wcf}
{\bf ARA} Collaboration, P.~Allison et~al., {\it {Constraints on the diffuse
  flux of ultrahigh energy neutrinos from four years of Askaryan Radio Array
  data in two stations}},  {\em Phys. Rev. D} {\bf 102} (2020), no.~4 043021,
  [\href{http://arxiv.org/abs/1912.00987}{{\tt arXiv:1912.00987}}].

\bibitem{ARIANNA:2019scz}
{\bf ARIANNA} Collaboration, A.~Anker et~al., {\it {Targeting ultra-high energy
  neutrinos with the ARIANNA experiment}},  {\em Adv. Space Res.} {\bf 64}
  (2019) 2595--2609, [\href{http://arxiv.org/abs/1903.01609}{{\tt
  arXiv:1903.01609}}].

\bibitem{GRAND:2018iaj}
{\bf GRAND} Collaboration, J.~\'Alvarez-Mu\~niz et~al., {\it {The Giant Radio
  Array for Neutrino Detection (GRAND): Science and Design}},  {\em Sci. China
  Phys. Mech. Astron.} {\bf 63} (2020), no.~1 219501,
  [\href{http://arxiv.org/abs/1810.09994}{{\tt arXiv:1810.09994}}].

\bibitem{IceCube-Gen2:2021rkf}
{\bf IceCube-Gen2} Collaboration, R.~Abbasi et~al., {\it {Sensitivity studies
  for the IceCube-Gen2 radio array}},  {\em PoS} {\bf ICRC2021} (2021) 1183,
  [\href{http://arxiv.org/abs/2107.08910}{{\tt arXiv:2107.08910}}].

\bibitem{POEMMA:2020ykm}
{\bf POEMMA} Collaboration, A.~V. Olinto et~al., {\it {The POEMMA (Probe of
  Extreme Multi-Messenger Astrophysics) observatory}},  {\em JCAP} {\bf 06}
  (2021) 007, [\href{http://arxiv.org/abs/2012.07945}{{\tt arXiv:2012.07945}}].

\bibitem{IceCube-Gen2:2020qha}
{\bf IceCube-Gen2} Collaboration, M.~G. Aartsen et~al., {\it {IceCube-Gen2: the
  window to the extreme Universe}},  {\em J. Phys. G} {\bf 48} (2021), no.~6
  060501, [\href{http://arxiv.org/abs/2008.04323}{{\tt arXiv:2008.04323}}].

\bibitem{Connolly:2011vc}
A.~Connolly, R.~S. Thorne, and D.~Waters, {\it {Calculation of High Energy
  Neutrino-Nucleon Cross Sections and Uncertainties Using the MSTW Parton
  Distribution Functions and Implications for Future Experiments}},  {\em Phys.
  Rev. D} {\bf 83} (2011) 113009, [\href{http://arxiv.org/abs/1102.0691}{{\tt
  arXiv:1102.0691}}].

\bibitem{Arguelles:2015wba}
C.~A. Arg\"uelles, F.~Halzen, L.~Wille, M.~Kroll, and M.~H. Reno, {\it
  {High-energy behavior of photon, neutrino, and proton cross sections}},  {\em
  Phys. Rev. D} {\bf 92} (2015), no.~7 074040,
  [\href{http://arxiv.org/abs/1504.06639}{{\tt arXiv:1504.06639}}].

\bibitem{Garcia:2020jwr}
A.~Garcia, R.~Gauld, A.~Heijboer, and J.~Rojo, {\it {Complete predictions for
  high-energy neutrino propagation in matter}},  {\em JCAP} {\bf 09} (2020)
  025, [\href{http://arxiv.org/abs/2004.04756}{{\tt arXiv:2004.04756}}].

\bibitem{Plestid:2024bva}
R.~Plestid and B.~Zhou, {\it {Final state radiation from high and ultrahigh
  energy neutrino interactions}},  \href{http://arxiv.org/abs/2403.07984}{{\tt
  arXiv:2403.07984}}.

\bibitem{PierreAuger:2024flk}
{\bf Pierre Auger} Collaboration, A.~Abdul~Halim et~al., {\it {Inference of the
  Mass Composition of Cosmic Rays with Energies from 1018.5 to 1020\,\,eV Using
  the Pierre Auger Observatory and Deep Learning}},  {\em Phys. Rev. Lett.}
  {\bf 134} (2025), no.~2 021001, [\href{http://arxiv.org/abs/2406.06315}{{\tt
  arXiv:2406.06315}}].

\bibitem{TelescopeArray:2024buq}
{\bf Telescope Array} Collaboration, R.~U. Abbasi et~al., {\it {Mass
  composition of ultrahigh energy cosmic rays from distribution of their
  arrival directions with the Telescope Array}},  {\em Phys. Rev. D} {\bf 110}
  (2024), no.~2 022006, [\href{http://arxiv.org/abs/2406.19286}{{\tt
  arXiv:2406.19286}}].

\bibitem{Romero-Wolf:2017xqe}
A.~Romero-Wolf and M.~Ave, {\it {Bayesian Inference Constraints on
  Astrophysical Production of Ultra-high Energy Cosmic Rays and Cosmogenic
  Neutrino Flux Predictions}},  {\em JCAP} {\bf 07} (2018) 025,
  [\href{http://arxiv.org/abs/1712.07290}{{\tt arXiv:1712.07290}}].

\bibitem{Heinze:2019jou}
J.~Heinze, A.~Fedynitch, D.~Boncioli, and W.~Winter, {\it {A new view on Auger
  data and cosmogenic neutrinos in light of different nuclear disintegration
  and air-shower models}},  {\em Astrophys. J.} {\bf 873} (2019), no.~1 88,
  [\href{http://arxiv.org/abs/1901.03338}{{\tt arXiv:1901.03338}}].

\bibitem{Ahlers:2017bV}
M.~Ahlers, P.~Denton, and M.~Rameez, {\it {Analyzing UHECR arrival directions
  through the Galactic magnetic field in view of the local universe as seen in
  2MRS}},  {\em PoS} {\bf ICRC2017} (2017) 282.

\bibitem{PierreAuger:2022atd}
{\bf Pierre Auger} Collaboration, A.~A. Halim et~al., {\it {Constraining the
  sources of ultra-high-energy cosmic rays across and above the ankle with the
  spectrum and composition data measured at the Pierre Auger Observatory}},
  {\em JCAP} {\bf 05} (2023) 024, [\href{http://arxiv.org/abs/2211.02857}{{\tt
  arXiv:2211.02857}}].

\bibitem{TelescopeArray:2024oux}
{\bf Telescope Array} Collaboration, R.~U. Abbasi et~al., {\it {Isotropy of
  Cosmic Rays beyond 1020\,\,eV Favors Their Heavy Mass Composition}},  {\em
  Phys. Rev. Lett.} {\bf 133} (2024), no.~4 041001,
  [\href{http://arxiv.org/abs/2406.19287}{{\tt arXiv:2406.19287}}].

\bibitem{Moller:2018isk}
K.~M\o{}ller, P.~B. Denton, and I.~Tamborra, {\it {Cosmogenic Neutrinos Through
  the GRAND Lens Unveil the Nature of Cosmic Accelerators}},  {\em JCAP} {\bf
  05} (2019) 047, [\href{http://arxiv.org/abs/1809.04866}{{\tt
  arXiv:1809.04866}}].

\bibitem{vanVliet:2019nse}
A.~van Vliet, R.~Alves~Batista, and J.~R. H\"orandel, {\it {Determining the
  fraction of cosmic-ray protons at ultrahigh energies with cosmogenic
  neutrinos}},  {\em Phys. Rev. D} {\bf 100} (2019), no.~2 021302,
  [\href{http://arxiv.org/abs/1901.01899}{{\tt arXiv:1901.01899}}].

\bibitem{Yuksel:2008cu}
H.~Yuksel, M.~D. Kistler, J.~F. Beacom, and A.~M. Hopkins, {\it {Revealing the
  High-Redshift Star Formation Rate with Gamma-Ray Bursts}},  {\em Astrophys.
  J. Lett.} {\bf 683} (2008) L5--L8,
  [\href{http://arxiv.org/abs/0804.4008}{{\tt arXiv:0804.4008}}].

\bibitem{Caccianiga:2001vu}
A.~Caccianiga, T.~Maccacaro, A.~Wolter, R.~Della~Ceca, and I.~M. Gioia, {\it
  {On the cosmological evolution of bl lacs}},  {\em Astrophys. J.} {\bf 566}
  (2002) 181, [\href{http://arxiv.org/abs/astro-ph/0110334}{{\tt
  astro-ph/0110334}}].

\bibitem{Ajello:2013lka}
M.~Ajello et~al., {\it {The Cosmic Evolution of Fermi BL Lacertae Objects}},
  {\em Astrophys. J.} {\bf 780} (2014) 73,
  [\href{http://arxiv.org/abs/1310.0006}{{\tt arXiv:1310.0006}}].

\bibitem{AlvesBatista:2016vpy}
R.~Alves~Batista, A.~Dundovic, M.~Erdmann, K.-H. Kampert, D.~Kuempel,
  G.~M\"uller, G.~Sigl, A.~van Vliet, D.~Walz, and T.~Winchen, {\it {CRPropa 3
  - a Public Astrophysical Simulation Framework for Propagating
  Extraterrestrial Ultra-High Energy Particles}},  {\em JCAP} {\bf 05} (2016)
  038, [\href{http://arxiv.org/abs/1603.07142}{{\tt arXiv:1603.07142}}].

\bibitem{Wittkowski:2018giy}
D.~Wittkowski and K.-H. Kampert, {\it {On the flux of high-energy cosmogenic
  neutrinos and the influence of the extragalactic magnetic field}},  {\em Mon.
  Not. Roy. Astron. Soc.} {\bf 488} (2019), no.~1 L119--L122,
  [\href{http://arxiv.org/abs/1810.03769}{{\tt arXiv:1810.03769}}].

\bibitem{2012MNRAS.422.3189G}
R.~C. {Gilmore}, R.~S. {Somerville}, J.~R. {Primack}, and A.~{Dom{\'\i}nguez},
  {\it {Semi-analytic modelling of the extragalactic background light and
  consequences for extragalactic gamma-ray spectra}},  {\em MNRAS} {\bf 422}
  (June, 2012) 3189--3207, [\href{http://arxiv.org/abs/1104.0671}{{\tt
  arXiv:1104.0671}}].

\bibitem{AlvesBatista:2019rhs}
R.~Alves~Batista, D.~Boncioli, A.~di~Matteo, and A.~van Vliet, {\it {Secondary
  neutrino and gamma-ray fluxes from SimProp and CRPropa}},  {\em JCAP} {\bf
  05} (2019) 006, [\href{http://arxiv.org/abs/1901.01244}{{\tt
  arXiv:1901.01244}}].

\bibitem{PierreAuger:2020qqz}
{\bf Pierre Auger} Collaboration, A.~Aab et~al., {\it {Measurement of the
  cosmic-ray energy spectrum above $2.5{\times} 10^{18}$ eV using the Pierre
  Auger Observatory}},  {\em Phys. Rev. D} {\bf 102} (2020), no.~6 062005,
  [\href{http://arxiv.org/abs/2008.06486}{{\tt arXiv:2008.06486}}].

\bibitem{FENU20233531}
F.~Fenu, {\it The cosmic ray energy spectrum measured with the pierre auger
  observatory},  {\em Advances in Space Research} {\bf 72} (2023), no.~8
  3531--3537.

\bibitem{SajjadAthar:2021prg}
M.~Sajjad~Athar et~al., {\it {Status and perspectives of neutrino physics}},
  {\em Prog. Part. Nucl. Phys.} {\bf 124} (2022) 103947,
  [\href{http://arxiv.org/abs/2111.07586}{{\tt arXiv:2111.07586}}].

\bibitem{Gandhi:1998ri}
R.~Gandhi, C.~Quigg, M.~H. Reno, and I.~Sarcevic, {\it {Neutrino interactions
  at ultrahigh-energies}},  {\em Phys. Rev. D} {\bf 58} (1998) 093009,
  [\href{http://arxiv.org/abs/hep-ph/9807264}{{\tt hep-ph/9807264}}].

\bibitem{Askaryan:1962}
G.~A. Askaryan, {\it {Excess negative charge of an electron-photon shower and
  its coherent radio emission}},  {\em Sov. Phys. JETP} {\bf 14} (1962) 441.

\bibitem{Askaryan:1965}
G.~A. Askaryan, {\it {Coherent Radio Emission from Cosmic Showers in Air and in
  Dense Media}},  {\em Sov. Phys. JETP} {\bf 48} (1965) 988.

\bibitem{Testagrossa:2023ukh}
F.~Testagrossa, D.~F.~G. Fiorillo, and M.~Bustamante, {\it {Two-detector flavor
  sensitivity to ultrahigh-energy cosmic neutrinos}},  {\em Phys. Rev. D} {\bf
  110} (2024), no.~8 083026, [\href{http://arxiv.org/abs/2310.12215}{{\tt
  arXiv:2310.12215}}].

\bibitem{Beacom:2004yd}
J.~F. Beacom, N.~F. Bell, and S.~Dodelson, {\it {Neutrinoless universe}},  {\em
  Phys. Rev. Lett.} {\bf 93} (2004) 121302,
  [\href{http://arxiv.org/abs/astro-ph/0404585}{{\tt astro-ph/0404585}}].

\bibitem{Fiorillo:2020jvy}
D.~F.~G. Fiorillo, G.~Miele, S.~Morisi, and N.~Saviano, {\it {Cosmogenic
  neutrino fluxes under the effect of active-sterile secret interactions}},
  {\em Phys. Rev. D} {\bf 101} (2020), no.~8 083024,
  [\href{http://arxiv.org/abs/2002.10125}{{\tt arXiv:2002.10125}}].

\bibitem{Blum:2014ewa}
K.~Blum, A.~Hook, and K.~Murase, {\it {High energy neutrino telescopes as a
  probe of the neutrino mass mechanism}},
  \href{http://arxiv.org/abs/1408.3799}{{\tt arXiv:1408.3799}}.

\bibitem{Berryman:2018ogk}
J.~M. Berryman, A.~De~Gouv\^ea, K.~J. Kelly, and Y.~Zhang, {\it
  {Lepton-Number-Charged Scalars and Neutrino Beamstrahlung}},  {\em Phys. Rev.
  D} {\bf 97} (2018), no.~7 075030,
  [\href{http://arxiv.org/abs/1802.00009}{{\tt arXiv:1802.00009}}].

\bibitem{Kelly:2020pcy}
K.~J. Kelly, M.~Sen, W.~Tangarife, and Y.~Zhang, {\it {Origin of sterile
  neutrino dark matter via secret neutrino interactions with vector bosons}},
  {\em Phys. Rev. D} {\bf 101} (2020), no.~11 115031,
  [\href{http://arxiv.org/abs/2005.03681}{{\tt arXiv:2005.03681}}].

\bibitem{Berryman:2022hds}
J.~M. Berryman et~al., {\it {Neutrino self-interactions: A white paper}},  {\em
  Phys. Dark Univ.} {\bf 42} (2023) 101267,
  [\href{http://arxiv.org/abs/2203.01955}{{\tt arXiv:2203.01955}}].

\bibitem{Esteban:2021tub}
I.~Esteban, S.~Pandey, V.~Brdar, and J.~F. Beacom, {\it {Probing secret
  interactions of astrophysical neutrinos in the high-statistics era}},  {\em
  Phys. Rev. D} {\bf 104} (2021), no.~12 123014,
  [\href{http://arxiv.org/abs/2107.13568}{{\tt arXiv:2107.13568}}].

\bibitem{Wang:2025qap}
I.~R. Wang, X.-J. Xu, and B.~Zhou, {\it {Widen the Resonance: Probing a New
  Regime of Neutrino Self-Interactions with Astrophysical Neutrinos}},
  \href{http://arxiv.org/abs/2501.07624}{{\tt arXiv:2501.07624}}.

\bibitem{Blinov:2019gcj}
N.~Blinov, K.~J. Kelly, G.~Z. Krnjaic, and S.~D. McDermott, {\it {Constraining
  the Self-Interacting Neutrino Interpretation of the Hubble Tension}},  {\em
  Phys. Rev. Lett.} {\bf 123} (2019), no.~19 191102,
  [\href{http://arxiv.org/abs/1905.02727}{{\tt arXiv:1905.02727}}].

\bibitem{DiValentino:2024xsv}
E.~Di~Valentino, S.~Gariazzo, and O.~Mena, {\it {Neutrinos in Cosmology}},
  \href{http://arxiv.org/abs/2404.19322}{{\tt arXiv:2404.19322}}.

\bibitem{Ng:2014pca}
K.~C.~Y. Ng and J.~F. Beacom, {\it {Cosmic neutrino cascades from secret
  neutrino interactions}},  {\em Phys. Rev. D} {\bf 90} (2014), no.~6 065035,
  [\href{http://arxiv.org/abs/1404.2288}{{\tt arXiv:1404.2288}}]. [Erratum:
  Phys.Rev.D 90, 089904 (2014)].

\bibitem{Creque-Sarbinowski:2020qhz}
C.~Creque-Sarbinowski, J.~Hyde, and M.~Kamionkowski, {\it {Resonant neutrino
  self-interactions}},  {\em Phys. Rev. D} {\bf 103} (2021), no.~2 023527,
  [\href{http://arxiv.org/abs/2005.05332}{{\tt arXiv:2005.05332}}].

\bibitem{Wolfenstein:1981kw}
L.~Wolfenstein, {\it {Different Varieties of Massive Dirac Neutrinos}},  {\em
  Nucl. Phys. B} {\bf 186} (1981) 147--152.

\bibitem{Petcov:1982ya}
S.~T. Petcov, {\it {On Pseudodirac Neutrinos, Neutrino Oscillations and
  Neutrinoless Double beta Decay}},  {\em Phys. Lett. B} {\bf 110} (1982)
  245--249.

\bibitem{Valle:1983dk}
J.~W.~F. Valle and M.~Singer, {\it {Lepton Number Violation With Quasi Dirac
  Neutrinos}},  {\em Phys. Rev. D} {\bf 28} (1983) 540.

\bibitem{Kallosh:1995hi}
R.~Kallosh, A.~D. Linde, D.~A. Linde, and L.~Susskind, {\it {Gravity and global
  symmetries}},  {\em Phys. Rev. D} {\bf 52} (1995) 912--935,
  [\href{http://arxiv.org/abs/hep-th/9502069}{{\tt hep-th/9502069}}].

\bibitem{Chang:1999pb}
D.~Chang and O.~C.~W. Kong, {\it {Pseudo-Dirac neutrinos}},  {\em Phys. Lett.
  B} {\bf 477} (2000) 416--423,
  [\href{http://arxiv.org/abs/hep-ph/9912268}{{\tt hep-ph/9912268}}].

\bibitem{Joshipura:2000ts}
A.~S. Joshipura and S.~D. Rindani, {\it {Phenomenology of pseudoDirac
  neutrinos}},  {\em Phys. Lett. B} {\bf 494} (2000) 114--123,
  [\href{http://arxiv.org/abs/hep-ph/0007334}{{\tt hep-ph/0007334}}].

\bibitem{Lindner:2001hr}
M.~Lindner, T.~Ohlsson, and G.~Seidl, {\it {Seesaw mechanisms for Dirac and
  Majorana neutrino masses}},  {\em Phys. Rev. D} {\bf 65} (2002) 053014,
  [\href{http://arxiv.org/abs/hep-ph/0109264}{{\tt hep-ph/0109264}}].

\bibitem{Berezinsky:2002fa}
V.~Berezinsky, M.~Narayan, and F.~Vissani, {\it {Mirror model for sterile
  neutrinos}},  {\em Nucl. Phys. B} {\bf 658} (2003) 254--280,
  [\href{http://arxiv.org/abs/hep-ph/0210204}{{\tt hep-ph/0210204}}].

\bibitem{Ma:2014qra}
E.~Ma and R.~Srivastava, {\it {Dirac or inverse seesaw neutrino masses with
  $B-L$ gauge symmetry and $S_3$ flavor symmetry}},  {\em Phys. Lett. B} {\bf
  741} (2015) 217--222, [\href{http://arxiv.org/abs/1411.5042}{{\tt
  arXiv:1411.5042}}].

\bibitem{Valle:2016kyz}
J.~W.~F. Valle and C.~A. Vaquera-Araujo, {\it {Dynamical seesaw mechanism for
  Dirac neutrinos}},  {\em Phys. Lett. B} {\bf 755} (2016) 363--366,
  [\href{http://arxiv.org/abs/1601.05237}{{\tt arXiv:1601.05237}}].

\bibitem{Ahn:2016hhq}
Y.~H. Ahn, S.~K. Kang, and C.~S. Kim, {\it {A Model for Pseudo-Dirac Neutrinos:
  Leptogenesis and Ultra-High Energy Neutrinos}},  {\em JHEP} {\bf 10} (2016)
  092, [\href{http://arxiv.org/abs/1602.05276}{{\tt arXiv:1602.05276}}].

\bibitem{Babu:2022ikf}
K.~S. Babu, X.-G. He, M.~Su, and A.~Thapa, {\it {Naturally light Dirac and
  pseudo-Dirac neutrinos from left-right symmetry}},  {\em JHEP} {\bf 08}
  (2022) 140, [\href{http://arxiv.org/abs/2205.09127}{{\tt arXiv:2205.09127}}].

\bibitem{Kobayashi:2000md}
M.~Kobayashi and C.~S. Lim, {\it {Pseudo Dirac scenario for neutrino
  oscillations}},  {\em Phys. Rev. D} {\bf 64} (2001) 013003,
  [\href{http://arxiv.org/abs/hep-ph/0012266}{{\tt hep-ph/0012266}}].

\bibitem{Martinez-Soler:2021unz}
I.~Martinez-Soler, Y.~F. Perez-Gonzalez, and M.~Sen, {\it {Signs of
  pseudo-Dirac neutrinos in SN1987A data}},  {\em Phys. Rev. D} {\bf 105}
  (2022), no.~9 095019, [\href{http://arxiv.org/abs/2105.12736}{{\tt
  arXiv:2105.12736}}].

\bibitem{Franklin:2023diy}
J.~Franklin, Y.~F. Perez-Gonzalez, and J.~Turner, {\it {JUNO as a probe of the
  pseudo-Dirac nature using solar neutrinos}},  {\em Phys. Rev. D} {\bf 108}
  (2023), no.~3 035010, [\href{http://arxiv.org/abs/2304.05418}{{\tt
  arXiv:2304.05418}}].

\bibitem{Carloni:2022cqz}
K.~Carloni, I.~Mart\'\i{}nez-Soler, C.~A. Arguelles, K.~S. Babu, and P.~S.~B.
  Dev, {\it {Probing pseudo-Dirac neutrinos with astrophysical sources at
  IceCube}},  {\em Phys. Rev. D} {\bf 109} (2024) L051702,
  [\href{http://arxiv.org/abs/2212.00737}{{\tt arXiv:2212.00737}}].

\bibitem{Dev:2024yrg}
P.~S.~B. Dev, P.~A.~N. Machado, and I.~Martinez-Soler, {\it {Pseudo-Dirac
  Neutrinos and Relic Neutrino Matter Effect on the High-energy Neutrino Flavor
  Composition}},  \href{http://arxiv.org/abs/2406.18507}{{\tt
  arXiv:2406.18507}}.

\bibitem{Esteban:2024eli}
I.~Esteban, M.~C. Gonzalez-Garcia, M.~Maltoni, I.~Martinez-Soler, J.~a.~P.
  Pinheiro, and T.~Schwetz, {\it {NuFit-6.0: Updated global analysis of
  three-flavor neutrino oscillations}},
  \href{http://arxiv.org/abs/2410.05380}{{\tt arXiv:2410.05380}}.

\bibitem{Brdar:2017kbt}
V.~Brdar, J.~Kopp, J.~Liu, P.~Prass, and X.-P. Wang, {\it {Fuzzy dark matter
  and nonstandard neutrino interactions}},  {\em Phys. Rev. D} {\bf 97} (2018),
  no.~4 043001, [\href{http://arxiv.org/abs/1705.09455}{{\tt
  arXiv:1705.09455}}].

\bibitem{Abbott:1982af}
L.~F. Abbott and P.~Sikivie, {\it {A Cosmological Bound on the Invisible
  Axion}},  {\em Phys. Lett. B} {\bf 120} (1983) 133--136.

\bibitem{Dine:1982ah}
M.~Dine and W.~Fischler, {\it {The Not So Harmless Axion}},  {\em Phys. Lett.
  B} {\bf 120} (1983) 137--141.

\bibitem{Preskill:1982cy}
J.~Preskill, M.~B. Wise, and F.~Wilczek, {\it {Cosmology of the Invisible
  Axion}},  {\em Phys. Lett. B} {\bf 120} (1983) 127--132.

\bibitem{Chun:2021ief}
E.~J. Chun, {\it {Neutrino Transition in Dark Matter}},
  \href{http://arxiv.org/abs/2112.05057}{{\tt arXiv:2112.05057}}.

\bibitem{Berlin:2016woy}
A.~Berlin, {\it {Neutrino Oscillations as a Probe of Light Scalar Dark
  Matter}},  {\em Phys. Rev. Lett.} {\bf 117} (2016), no.~23 231801,
  [\href{http://arxiv.org/abs/1608.01307}{{\tt arXiv:1608.01307}}].

\bibitem{Krnjaic:2017zlz}
G.~Krnjaic, P.~A.~N. Machado, and L.~Necib, {\it {Distorted neutrino
  oscillations from time varying cosmic fields}},  {\em Phys. Rev. D} {\bf 97}
  (2018), no.~7 075017, [\href{http://arxiv.org/abs/1705.06740}{{\tt
  arXiv:1705.06740}}].

\bibitem{Capozzi:2018bps}
F.~Capozzi, I.~M. Shoemaker, and L.~Vecchi, {\it {Neutrino Oscillations in Dark
  Backgrounds}},  {\em JCAP} {\bf 07} (2018) 004,
  [\href{http://arxiv.org/abs/1804.05117}{{\tt arXiv:1804.05117}}].

\bibitem{Dev:2020kgz}
A.~Dev, P.~A.~N. Machado, and P.~Mart\'\i{}nez-Mirav\'e, {\it {Signatures of
  ultralight dark matter in neutrino oscillation experiments}},  {\em JHEP}
  {\bf 01} (2021) 094, [\href{http://arxiv.org/abs/2007.03590}{{\tt
  arXiv:2007.03590}}].

\bibitem{Losada:2021bxx}
M.~Losada, Y.~Nir, G.~Perez, and Y.~Shpilman, {\it {Probing scalar dark matter
  oscillations with neutrino oscillations}},  {\em JHEP} {\bf 04} (2022) 030,
  [\href{http://arxiv.org/abs/2107.10865}{{\tt arXiv:2107.10865}}].

\bibitem{Dev:2022bae}
A.~Dev, G.~Krnjaic, P.~Machado, and H.~Ramani, {\it {Constraining feeble
  neutrino interactions with ultralight dark matter}},  {\em Phys. Rev. D} {\bf
  107} (2023), no.~3 035006, [\href{http://arxiv.org/abs/2205.06821}{{\tt
  arXiv:2205.06821}}].

\bibitem{Davoudiasl:2023uiq}
H.~Davoudiasl and P.~B. Denton, {\it {Sterile neutrino shape shifting caused by
  dark matter}},  {\em Phys. Rev. D} {\bf 108} (2023), no.~3 035013,
  [\href{http://arxiv.org/abs/2301.09651}{{\tt arXiv:2301.09651}}].

\bibitem{Gherghetta:2023myo}
T.~Gherghetta and A.~Shkerin, {\it {Probing a local dark matter halo with
  neutrino oscillations}},  {\em Phys. Rev. D} {\bf 108} (2023), no.~9 095009,
  [\href{http://arxiv.org/abs/2305.06441}{{\tt arXiv:2305.06441}}].

\bibitem{Sen:2023uga}
M.~Sen and A.~Y. Smirnov, {\it {Refractive neutrino masses, ultralight dark
  matter and cosmology}},  {\em JCAP} {\bf 01} (2024) 040,
  [\href{http://arxiv.org/abs/2306.15718}{{\tt arXiv:2306.15718}}].

\bibitem{Kamiokande-II:1987idp}
{\bf Kamiokande-II} Collaboration, K.~Hirata et~al., {\it {Observation of a
  Neutrino Burst from the Supernova SN 1987a}},  {\em Phys. Rev. Lett.} {\bf
  58} (1987) 1490--1493.

\bibitem{Hirata:1988ad}
K.~S. Hirata et~al., {\it {Observation in the Kamiokande-II Detector of the
  Neutrino Burst from Supernova SN 1987a}},  {\em Phys. Rev. D} {\bf 38} (1988)
  448--458.

\bibitem{Bionta:1987qt}
R.~M. Bionta et~al., {\it {Observation of a Neutrino Burst in Coincidence with
  Supernova SN 1987a in the Large Magellanic Cloud}},  {\em Phys. Rev. Lett.}
  {\bf 58} (1987) 1494.

\bibitem{Alekseev:1988gp}
E.~N. Alekseev, L.~N. Alekseeva, I.~V. Krivosheina, and V.~I. Volchenko, {\it
  {Detection of the Neutrino Signal From {SN1987A} in the {LMC} Using the Inr
  Baksan Underground Scintillation Telescope}},  {\em Phys. Lett. B} {\bf 205}
  (1988) 209--214.

\bibitem{IceCube:2018dnn}
{\bf IceCube, Fermi-LAT, MAGIC, AGILE, ASAS-SN, HAWC, H.E.S.S., INTEGRAL,
  Kanata, Kiso, Kapteyn, Liverpool Telescope, Subaru, Swift NuSTAR, VERITAS,
  VLA/17B-403} Collaboration, M.~G. Aartsen et~al., {\it {Multimessenger
  observations of a flaring blazar coincident with high-energy neutrino
  IceCube-170922A}},  {\em Science} {\bf 361} (2018), no.~6398 eaat1378,
  [\href{http://arxiv.org/abs/1807.08816}{{\tt arXiv:1807.08816}}].

\bibitem{Bertoni:2014mva}
B.~Bertoni, S.~Ipek, D.~McKeen, and A.~E. Nelson, {\it {Constraints and
  consequences of reducing small scale structure via large dark matter-neutrino
  interactions}},  {\em JHEP} {\bf 04} (2015) 170,
  [\href{http://arxiv.org/abs/1412.3113}{{\tt arXiv:1412.3113}}].

\bibitem{Wilkinson:2014ksa}
R.~J. Wilkinson, C.~Boehm, and J.~Lesgourgues, {\it {Constraining Dark
  Matter-Neutrino Interactions using the CMB and Large-Scale Structure}},  {\em
  JCAP} {\bf 05} (2014) 011, [\href{http://arxiv.org/abs/1401.7597}{{\tt
  arXiv:1401.7597}}].

\bibitem{Escudero:2018thh}
M.~Escudero, L.~Lopez-Honorez, O.~Mena, S.~Palomares-Ruiz, and
  P.~Villanueva-Domingo, {\it {A fresh look into the interacting dark matter
  scenario}},  {\em JCAP} {\bf 06} (2018) 007,
  [\href{http://arxiv.org/abs/1803.08427}{{\tt arXiv:1803.08427}}].

\bibitem{Brax:2023tvn}
P.~Brax, C.~van~de Bruck, E.~Di~Valentino, W.~Giar\`e, and S.~Trojanowski, {\it
  {Extended analysis of neutrino-dark matter interactions with small-scale CMB
  experiments}},  {\em Phys. Dark Univ.} {\bf 42} (2023) 101321,
  [\href{http://arxiv.org/abs/2305.01383}{{\tt arXiv:2305.01383}}].

\bibitem{Murase:2007yt}
K.~Murase, {\it {High energy neutrino early afterglows gamma-ray bursts
  revisited}},  {\em Phys. Rev. D} {\bf 76} (2007) 123001,
  [\href{http://arxiv.org/abs/0707.1140}{{\tt arXiv:0707.1140}}].

\bibitem{Fang:2017zjf}
K.~Fang and K.~Murase, {\it {Linking High-Energy Cosmic Particles by Black Hole
  Jets Embedded in Large-Scale Structures}},  {\em Nature Phys.} {\bf 14}
  (2018), no.~4 396--398, [\href{http://arxiv.org/abs/1704.00015}{{\tt
  arXiv:1704.00015}}].

\bibitem{Alves:2025xul}
G.~F.~S. Alves, M.~Hostert, and M.~Pospelov, {\it {Neutron portal to
  ultra-high-energy neutrinos}},  \href{http://arxiv.org/abs/2503.14419}{{\tt
  arXiv:2503.14419}}.

\bibitem{Righi:2020ufi}
C.~Righi, A.~Palladino, F.~Tavecchio, and F.~Vissani, {\it {EeV astrophysical
  neutrinos from flat spectrum radio quasars}},  {\em Astron. Astrophys.} {\bf
  642} (2020) A92, [\href{http://arxiv.org/abs/2003.08701}{{\tt
  arXiv:2003.08701}}].

\bibitem{Rodrigues:2020pli}
X.~Rodrigues, J.~Heinze, A.~Palladino, A.~van Vliet, and W.~Winter, {\it
  {Active Galactic Nuclei Jets as the Origin of Ultrahigh-Energy Cosmic Rays
  and Perspectives for the Detection of Astrophysical Source Neutrinos at EeV
  Energies}},  {\em Phys. Rev. Lett.} {\bf 126} (2021), no.~19 191101,
  [\href{http://arxiv.org/abs/2003.08392}{{\tt arXiv:2003.08392}}].

\bibitem{IceCube:2025ezc}
{\bf IceCube} Collaboration, R.~Abbasi et~al., {\it {A search for
  extremely-high-energy neutrinos and first constraints on the
  ultra-high-energy cosmic-ray proton fraction with IceCube}},
  \href{http://arxiv.org/abs/2502.01963}{{\tt arXiv:2502.01963}}.

\bibitem{Li:2025tqf}
S.~W. Li, P.~Machado, D.~Naredo-Tuero, and T.~Schwemberger, {\it {Clash of the
  Titans: ultra-high energy KM3NeT event versus IceCube data}},
  \href{http://arxiv.org/abs/2502.04508}{{\tt arXiv:2502.04508}}.

\bibitem{IceCube:2018fhm}
{\bf IceCube} Collaboration, M.~G. Aartsen et~al., {\it {Differential limit on
  the extremely-high-energy cosmic neutrino flux in the presence of
  astrophysical background from nine years of IceCube data}},  {\em Phys. Rev.
  D} {\bf 98} (2018), no.~6 062003,
  [\href{http://arxiv.org/abs/1807.01820}{{\tt arXiv:1807.01820}}].

\bibitem{Pasquini:2015fjv}
P.~S. Pasquini and O.~L.~G. Peres, {\it {Bounds on Neutrino-Scalar Yukawa
  Coupling}},  {\em Phys. Rev. D} {\bf 93} (2016), no.~5 053007,
  [\href{http://arxiv.org/abs/1511.01811}{{\tt arXiv:1511.01811}}]. [Erratum:
  Phys.Rev.D 93, 079902 (2016)].

\bibitem{deGouvea:2019qaz}
A.~de~Gouv\^ea, P.~S.~B. Dev, B.~Dutta, T.~Ghosh, T.~Han, and Y.~Zhang, {\it
  {Leptonic Scalars at the LHC}},  {\em JHEP} {\bf 07} (2020) 142,
  [\href{http://arxiv.org/abs/1910.01132}{{\tt arXiv:1910.01132}}].

\bibitem{Lessa:2007up}
A.~P. Lessa and O.~L.~G. Peres, {\it {Revising limits on neutrino-Majoron
  couplings}},  {\em Phys. Rev. D} {\bf 75} (2007) 094001,
  [\href{http://arxiv.org/abs/hep-ph/0701068}{{\tt hep-ph/0701068}}].

\bibitem{KamLAND-Zen:2012uen}
{\bf KamLAND-Zen} Collaboration, A.~Gando et~al., {\it {Limits on
  Majoron-emitting double-beta decays of Xe-136 in the KamLAND-Zen
  experiment}},  {\em Phys. Rev. C} {\bf 86} (2012) 021601,
  [\href{http://arxiv.org/abs/1205.6372}{{\tt arXiv:1205.6372}}].

\bibitem{Kharusi:2021jez}
S.~A. Kharusi et~al., {\it {Search for Majoron-emitting modes of $^{136}$Xe
  double beta decay with the complete EXO-200 dataset}},  {\em Phys. Rev. D}
  {\bf 104} (2021), no.~11 112002, [\href{http://arxiv.org/abs/2109.01327}{{\tt
  arXiv:2109.01327}}].

\bibitem{Blum:2018ljv}
K.~Blum, Y.~Nir, and M.~Shavit, {\it {Neutrinoless double-beta decay with
  massive scalar emission}},  {\em Phys. Lett. B} {\bf 785} (2018) 354--361,
  [\href{http://arxiv.org/abs/1802.08019}{{\tt arXiv:1802.08019}}].

\bibitem{Farzan:2018gtr}
Y.~Farzan, M.~Lindner, W.~Rodejohann, and X.-J. Xu, {\it {Probing neutrino
  coupling to a light scalar with coherent neutrino scattering}},  {\em JHEP}
  {\bf 05} (2018) 066, [\href{http://arxiv.org/abs/1802.05171}{{\tt
  arXiv:1802.05171}}].

\bibitem{AtzoriCorona:2022moj}
M.~Atzori~Corona, M.~Cadeddu, N.~Cargioli, F.~Dordei, C.~Giunti, Y.~F. Li,
  E.~Picciau, C.~A. Ternes, and Y.~Y. Zhang, {\it {Probing light mediators and
  (g \ensuremath{-} 2)$_{\mu}$ through detection of coherent elastic neutrino
  nucleus scattering at COHERENT}},  {\em JHEP} {\bf 05} (2022) 109,
  [\href{http://arxiv.org/abs/2202.11002}{{\tt arXiv:2202.11002}}].

\bibitem{Heurtier:2016otg}
L.~Heurtier and Y.~Zhang, {\it {Supernova Constraints on Massive (Pseudo)Scalar
  Coupling to Neutrinos}},  {\em JCAP} {\bf 02} (2017) 042,
  [\href{http://arxiv.org/abs/1609.05882}{{\tt arXiv:1609.05882}}].

\bibitem{Fiorillo:2022cdq}
D.~F.~G. Fiorillo, G.~G. Raffelt, and E.~Vitagliano, {\it {Strong Supernova
  1987A Constraints on Bosons Decaying to Neutrinos}},  {\em Phys. Rev. Lett.}
  {\bf 131} (2023), no.~2 021001, [\href{http://arxiv.org/abs/2209.11773}{{\tt
  arXiv:2209.11773}}].

\bibitem{Chen:2022kal}
Y.-M. Chen, M.~Sen, W.~Tangarife, D.~Tuckler, and Y.~Zhang, {\it {Core-collapse
  supernova constraint on the origin of sterile neutrino dark matter via
  neutrino self-interactions}},  {\em JCAP} {\bf 11} (2022) 014,
  [\href{http://arxiv.org/abs/2207.14300}{{\tt arXiv:2207.14300}}].

\bibitem{Huang:2017egl}
G.-y. Huang, T.~Ohlsson, and S.~Zhou, {\it {Observational Constraints on Secret
  Neutrino Interactions from Big Bang Nucleosynthesis}},  {\em Phys. Rev. D}
  {\bf 97} (2018), no.~7 075009, [\href{http://arxiv.org/abs/1712.04792}{{\tt
  arXiv:1712.04792}}].

\bibitem{Ioka:2014kca}
K.~Ioka and K.~Murase, {\it {IceCube PeV\textendash{}EeV neutrinos and secret
  interactions of neutrinos}},  {\em PTEP} {\bf 2014} (2014), no.~6 061E01,
  [\href{http://arxiv.org/abs/1404.2279}{{\tt arXiv:1404.2279}}].

\bibitem{Murase:2019xqi}
K.~Murase and I.~M. Shoemaker, {\it {Neutrino Echoes from Multimessenger
  Transient Sources}},  {\em Phys. Rev. Lett.} {\bf 123} (2019), no.~24 241102,
  [\href{http://arxiv.org/abs/1903.08607}{{\tt arXiv:1903.08607}}].

\bibitem{Kreisch_2020}
C.~D. Kreisch, F.-Y. Cyr-Racine, and O.~Doré, {\it Neutrino puzzle: Anomalies,
  interactions, and cosmological tensions},  {\em Physical Review D} {\bf 101}
  (June, 2020).

\bibitem{Poudou:2025qcx}
A.~Poudou, T.~Simon, T.~Montandon, E.~M. Teixeira, and V.~Poulin, {\it
  {Self-interacting neutrinos in light of recent CMB and LSS data}},
  \href{http://arxiv.org/abs/2503.10485}{{\tt arXiv:2503.10485}}.

\bibitem{Chen:2022zts}
Z.~Chen, J.~Liao, J.~Ling, and B.~Yue, {\it {Constraining super-light sterile
  neutrinos at Borexino and KamLAND}},  {\em JHEP} {\bf 09} (2022) 004,
  [\href{http://arxiv.org/abs/2205.07574}{{\tt arXiv:2205.07574}}].

\bibitem{Ansarifard:2022kvy}
S.~Ansarifard and Y.~Farzan, {\it {Revisiting pseudo-Dirac neutrino scenario
  after recent solar neutrino data}},  {\em Phys. Rev. D} {\bf 107} (2023),
  no.~7 075029, [\href{http://arxiv.org/abs/2211.09105}{{\tt
  arXiv:2211.09105}}].

\bibitem{Beacom:2003eu}
J.~F. Beacom, N.~F. Bell, D.~Hooper, J.~G. Learned, S.~Pakvasa, and T.~J.
  Weiler, {\it {PseudoDirac neutrinos: A Challenge for neutrino telescopes}},
  {\em Phys. Rev. Lett.} {\bf 92} (2004) 011101,
  [\href{http://arxiv.org/abs/hep-ph/0307151}{{\tt hep-ph/0307151}}].

\bibitem{Rink:2022nvw}
T.~Rink and M.~Sen, {\it {Constraints on pseudo-Dirac neutrinos using
  high-energy neutrinos from NGC 1068}},  {\em Phys. Lett. B} {\bf 851} (2024)
  138558, [\href{http://arxiv.org/abs/2211.16520}{{\tt arXiv:2211.16520}}].

\bibitem{Dixit:2024ldv}
K.~Dixit, L.~S. Miranda, and S.~Razzaque, {\it {Searching for Pseudo-Dirac
  neutrinos from Astrophysical sources in IceCube data}},
  \href{http://arxiv.org/abs/2406.06476}{{\tt arXiv:2406.06476}}.

\bibitem{Carloni:2025dhv}
K.~Carloni, Y.~Porto, C.~A. Arg\"uelles, P.~S.~B. Dev, and S.~Jana, {\it
  {Signatures of quasi-Dirac neutrinos in diffuse high-energy astrophysical
  neutrino data}},  \href{http://arxiv.org/abs/2503.19960}{{\tt
  arXiv:2503.19960}}.

\bibitem{Akita:2023yga}
K.~Akita and S.~Ando, {\it {Constraints on dark matter-neutrino scattering from
  the Milky-Way satellites and subhalo modeling for dark acoustic
  oscillations}},  {\em JCAP} {\bf 11} (2023) 037,
  [\href{http://arxiv.org/abs/2305.01913}{{\tt arXiv:2305.01913}}].

\bibitem{Ahlers:2009rf}
M.~Ahlers, L.~A. Anchordoqui, and S.~Sarkar, {\it {Neutrino diagnostics of
  ultra-high energy cosmic ray protons}},  {\em Phys. Rev. D} {\bf 79} (2009)
  083009, [\href{http://arxiv.org/abs/0902.3993}{{\tt arXiv:0902.3993}}].

\bibitem{Planck:2018vyg}
{\bf Planck} Collaboration, N.~Aghanim et~al., {\it {Planck 2018 results. VI.
  Cosmological parameters}},  {\em Astron. Astrophys.} {\bf 641} (2020) A6,
  [\href{http://arxiv.org/abs/1807.06209}{{\tt arXiv:1807.06209}}]. [Erratum:
  Astron.Astrophys. 652, C4 (2021)].

\end{thebibliography}\endgroup

\end{document}